\documentclass[11pt,a4paper]{article}
\usepackage{iftex}
\ifPDFTeX
  \pdfoutput=1
\fi
\usepackage{jheppub}
\usepackage[utf8]{inputenc}
\usepackage[T1]{fontenc}
\usepackage{amsmath,amssymb,mathtools,bm,mathrsfs}
\usepackage{slashed}
\usepackage{booktabs}
\usepackage{multirow}
\usepackage{xcolor}
\usepackage{microtype}
\usepackage{enumitem}
\usepackage{array}
\usepackage{placeins}
\usepackage[normalem]{ulem}

\ifPDFTeX
  \DeclareUnicodeCharacter{FF0C}{,}
\fi

\newcommand{\rF}{r_{\rm F}}
\newcommand{\rref}{r_{\rm F,ref}}
\newcommand{\MSbar}{\overline{\rm MS}}
\newcommand{\order}{\mathcal O}

\newcommand{\Mcal}{\mathcal M}
\newcommand{\Rcal}{\mathcal R}

\title{Gradient-flow ratio renormalization of nonlocal quasi operators in lattice QCD}

\author[a,b]{Haoyang Bai}
\author[c,d]{Junyuan Huang}
\author[c,d,\dagger]{Jun Hua}
\author[e]{Jia-Lu Zhang}
\author[e]{Mu-Hua Zhang}

\affiliation[a]{Institute of High Energy Physics, Chinese Academy of Sciences, Beijing 100049, China}
\affiliation[b]{University of Chinese Academy of Sciences, Beijing 100049, China}

\affiliation[c]{State Key Laboratory of Nuclear Physics and Technology, Institute of Quantum Matter, South China Normal University, Guangzhou 510006, China}
\affiliation[d]{Guangdong Basic Research Center of Excellence for Structure and Fundamental Interactions of Matter, Guangdong Provincial Key Laboratory of Nuclear Science, Guangzhou 510006, China}

\affiliation[e]{Tsung-Dao Lee Institute and School of Physics and Astronomy,
Shanghai Jiao Tong University, Shanghai 201210, China}

\emailAdd{$^\dagger$junhua@scnu.edu.cn}

\abstract{
We develop a gradient-flow ratio renormalization (GFRR) framework for nonlocal quasi
operators and determine the Wilson-line self-energy nonperturbatively
with high statistical precision.
Ratios at different flow radii cancel the common long-distance dependence in the small-flow-time regime, enabling a nonperturbative determination of the Wilson-line power-subtraction coefficient.
On two $N_f=2+1$ clover ensembles, we perform systematic consistency
tests of the self-energy determination using straight Wilson lines,
hadronic matrix elements, and the static potential, and examine its
lattice-spacing, momentum, and color-representation dependence.
We illustrate the numerical implementation at finite lattice spacing
using the pion quasi-distribution amplitude as an example.
By combining nonperturbative power subtraction, perturbative logarithmic
matching, and a short-distance determination of the finite residual mass,
GFRR provides a systematic prescription for renormalizing
nonlocal quasi operators into the $\MSbar$ and hybrid schemes.
}


\begin{document}
\maketitle
\flushbottom

\section{Introduction}
\label{sec:introduction}

Parton distribution functions (PDFs) and light-cone distribution amplitudes
(LCDAs) encode the longitudinal momentum structure of hadrons. Their
operator definitions involve lightlike separations, which preclude direct
evaluation in Euclidean lattice QCD. Large-momentum effective theory (LaMET)
enables their extraction from equal-time spatial correlations involving hadrons
with large momenta through perturbative matching
\cite{Ji:2013dva,Ji:2014gla,Ji:2020ect}.
These spatial correlations are matrix elements of nonlocal operators containing
a straight spatial Wilson line. A precise determination of PDFs and LCDAs
therefore requires both reliable statistical signals and controlled
renormalization of the ultraviolet divergences associated with these nonlocal
operators.

Gradient flow is well suited to addressing both requirements 
above. By smoothing
fields over a physical radius $\rF=\sqrt{8\tau}$, where $\tau$ is the flow
time, it suppresses ultraviolet fluctuations and can improve the numerical
signal of extended operators \cite{Luscher:2010iy,Shindler:2022tlx}.
In lattice QCD, gradient flow can be viewed as the continuous limit
of stout smearing \cite{Nagatsuka:2023jos}.
The flow also has a well-defined continuum formulation that provides a
systematic basis for operator renormalization.
After the usual renormalization of QCD and the additional normalization of
flowed fermion fields, local composite operators constructed from flowed fields
are ultraviolet finite at positive flow time
\cite{Luscher:2011bx,Luscher:2013cpa}. The short-flow-time expansion relates
these local flowed operators to renormalized operators in conventional
continuum schemes through perturbatively calculable coefficients, allowing
gradient flow to serve as an intermediate renormalization scheme.

Applications to local operators illustrate the scope of this approach.
The short-flow-time expansion has been used to construct renormalized
energy--momentum tensors and fermion bilinears
\cite{Makino:2014taa,Hieda:2016lly}.
For hadron structure, the gradient-flow formulation has opened a route to
higher Mellin moments of PDFs through local twist-two operators
\cite{Shindler:2023xpd}.
Recent lattice calculations have demonstrated this approach for the pion,
determining moment ratios up to $\langle x^5\rangle/\langle x\rangle$
\cite{Francis:2025rya,Francis:2025pgf}.

For spatially nonlocal operators, gradient flow has been used to define
smeared quasi-PDFs with a finite continuum limit at fixed positive flow time
\cite{Monahan:2016bvm}.
Subsequent studies derived one-loop matching relations between
flowed Wilson-line operators and their $\MSbar$ counterparts
\cite{Monahan:2017hpu,Brambilla:2023vwm,Zhang:2025npd}.
Recent work has extended the small-flow-time matching to
next-to-next-to-leading order for operators with the Wilson-line
self-energy subtracted \cite{Wang:2026toy}.
Numerical simulations also show benefits for related Wilson-line observables.
In calculations of the static force, gradient flow improves the
signal-to-noise ratio of Wilson loops with chromoelectric-field
insertions \cite{Brambilla:2023fsi}.
Studies of flowed Creutz ratios have found a weak lattice-spacing dependence
when the flow time is held fixed in physical units
\cite{Risch:2022vof}.
These matching results and numerical studies motivate the development
and testing of a nonperturbative renormalization procedure for flowed
quasi operators.

An important task in implementing such a procedure is to determine the
Wilson-line self-energy nonperturbatively and separate it from the physical
spatial dependence of the hadronic matrix element.
Existing mass-subtraction and self-renormalization approaches have provided ways
to address this problem in conventional lattice calculations \cite{Chen:2016fxx,LatticePartonLPC:2021gpi}.
Gradient flow offers additional information through a physical smoothing
radius that can be varied independently of the lattice spacing.
Forming ratios of the same observable at different flow radii on the same gauge
configurations allows the common long-distance dependence to cancel in the
small-flow-time regime, exposing the pure flow dependence of the self-energy.
Comparisons across ensembles at a common physical flow radius, in turn,
probe the remaining cutoff dependence.

In this work, we develop a gradient-flow ratio renormalization (GFRR) framework for
nonlocal quasi operators in Sec.~\ref{sec:theory} and numerically investigate
it on two $N_f=2+1$ clover ensembles, considering three cases: the pion
quasi-DA, the pion quasi-PDF, and the proton quasi-PDF.
In Sec.~\ref{sec:self-energy}, we determine the self-energy subtraction
coefficient from ratios of
Landau-gauge straight Wilson lines at different flow radii and compare this
determination with flow double ratios of hadronic matrix elements and a
gauge-invariant static-potential analysis.
Using the pion quasi-DA as an example, we demonstrate the coordinate-space
implementation of GFRR in Sec.~\ref{sec:step-by-step}, including conversion
to the $\MSbar$ scheme and
its incorporation into a hybrid scheme for subsequent LaMET matching.
GFRR combines a statistically precise nonperturbative
determination of the Wilson-line self-energy with separate prescriptions
for power subtraction, logarithmic renormalization, and the finite
residual mass.
Together, these steps provide a systematic procedure for renormalizing
nonlocal quasi operators.
The two lattice spacings allow us to monitor cutoff dependence, while a
controlled continuum extrapolation remains to be performed.
We summarize our findings in Sec.~\ref{sec:conclusion}.

\section{Theoretical framework}
\label{sec:theory}
We first review the gradient-flow formalism and the
renormalization of quasi-distributions, and then establish
the GFRR framework for flowed quasi-distributions.
\subsection{The gradient-flow formalism}
\label{subsec:gradient-flow}

In Euclidean spacetime, the gradient flow introduces an auxiliary
flow time $\tau\geq0$ and promotes the gauge and quark fields to
the flowed fields $B_\mu(\tau,x)$, $\chi(\tau,x)$, and
$\bar\chi(\tau,x)$. With the unflowed initial conditions
$B_\mu(0,x)=gA_\mu(x)$, $\chi(0,x)=\psi(x)$, and
$\bar\chi(0,x)=\bar\psi(x)$, the flow equations read
\begin{align}
\partial_\tau B_\mu
&=
D_\nu G_{\nu\mu}
+\kappa D_\mu\partial_\nu B_\nu,
\label{eq:flow_B}
\\
\partial_\tau\chi
&=
D_\mu D_\mu\chi
-\kappa(\partial_\mu B_\mu)\chi,
\label{eq:flow_chi}
\\
\partial_\tau\bar\chi
&=
\bar\chi\overleftarrow D_\mu\overleftarrow D_\mu
+\kappa\bar\chi\,\partial_\mu B_\mu.
\label{eq:flow_chibar}
\end{align}
Here we use the anti-Hermitian gauge-connection convention,
with the coupling absorbed into $B_\mu$, so that
$D_\mu=\partial_\mu+B_\mu$ on quarks,
$\overleftarrow D_\mu=\overleftarrow\partial_\mu-B_\mu$,
and
$G_{\mu\nu}=\partial_\mu B_\nu-\partial_\nu B_\mu
+[B_\mu,B_\nu]$.
The parameter $\kappa$ controls the damping of gauge modes
and does not affect gauge-invariant observables
\cite{Luscher:2011bx,Luscher:2013cpa}.
The flow introduces a smoothing radius $\rF=\sqrt{8\tau}$,
which can be varied independently of the lattice spacing.
For $a\ll\rF$, this provides a physical ultraviolet scale
at which the same operator can be studied on different
ensembles. 

The gradient flow can be formulated as a field theory in 4+1 dimensions, with $\tau$ as the additional coordinate.
The flow equations are implemented in the action by introducing
Lagrange-multiplier fields, with the QCD fields specified at
the boundary $\tau=0$. 
This formulation provides a perturbative framework for analyzing the
renormalization of flowed observables
\cite{Luscher:2011bx,Luscher:2013cpa}.
The renormalization analysis shows that, after the usual QCD
renormalization, no bulk counterterms are required at positive flow
time. 
For local composite operators constructed from
flowed fields, the only additional renormalization arises from
the flowed quark fields, which require a multiplicative
wave-function renormalization~\cite{Luscher:2013cpa}.
These properties allow renormalized flowed observables to be
defined in the continuum at fixed positive $\tau$, providing
a common intermediate scheme for lattice calculations and
continuum perturbation theory.

A standard prescription for eliminating the flowed-quark
wave-function factor is to use the ringed scheme
\cite{Makino:2014taa,Hieda:2016lly},
\begin{equation}
\mathring{\chi}(x,\tau)
=
\mathring Z_\chi^{1/2}\chi_0(x,\tau),
\qquad
\mathring{\bar\chi}(x,\tau)
=
\mathring Z_\chi^{1/2}\bar\chi_0(x,\tau),
\label{eq:ringed_fields}
\end{equation}
with
\begin{equation}
\mathring Z_\chi
=
\frac{-2N_c}
{(4\pi)^2\tau^2
\left.
\left\langle
\bar\chi_0(x,\tau)\,
\overleftrightarrow{\slashed D}\,
\chi_0(x,\tau)
\right\rangle
\right|_{m=0}},
\label{eq:ringed_Z}
\end{equation}
where $\overleftrightarrow{\slashed D}
=\gamma_\mu(D_\mu-\overleftarrow D_\mu)$,
and the normalization is written for a single quark flavor. This construction absorbs the
flowed-quark wave-function factor and specifies an intermediate
scheme for flowed composite operators that can be used in both
lattice calculations and continuum perturbation theory. 
Relating these finite flowed operators to operators in a conventional
continuum scheme requires the short-flow-time expansion (SFTX).
In the small-flow-time limit, the SFTX relates the ringed flowed quark fields to the
renormalized quark fields in the $\overline{\rm MS}$ scheme,
\begin{equation}
\begin{aligned}
\mathring{\chi}(x,\tau)
=
\mathring{\zeta}_\psi(\tau,\mu)\,
\psi^{\overline{\rm MS}}(x,\mu),\quad
\mathring{\bar\chi}(x,\tau)
=
\mathring{\zeta}_\psi(\tau,\mu)\,
\bar\psi^{\overline{\rm MS}}(x,\mu),
\end{aligned}
\label{eq:ringed_field_matching}
\end{equation}
where $\mu$ is the renormalization scale.

The SFTX requires the flow radius $r_F$ to be small compared
with the relevant physical length scales.
Small-flow-time matching has also been established for
nonlocal Wilson-line operators~\cite{Monahan:2017hpu,
Brambilla:2023vwm}.
Unlike local operators, these operators require an
additional mass renormalization of the Wilson line to
remove the linear divergence, together with a consistent
treatment of the associated UV renormalon, as discussed in
Sec.~\ref{subsec:flow-quasi-renormalization}.

\subsection{Quasi-distributions and their renormalization}
\label{subsec:quasi-renormalization}

In LaMET, light-cone partonic observables are extracted from
equal-time correlation functions in hadrons carrying a large
spatial momentum, which are accessible in Euclidean lattice
QCD~\cite{Ji:2013dva,Ji:2014gla,Ji:2020ect}.
At leading power in the large-momentum expansion, the
quasi-distributions and their light-cone counterparts share
the same infrared physics, while their ultraviolet differences
are accounted for by perturbatively calculable matching
coefficients. Corrections at finite hadron momentum are
organized in inverse powers of the hadron momentum.

The basic lattice observables are matrix elements of
spatially nonlocal operators containing a straight Wilson line.
For quark quasi-distributions, one considers the bilocal operator
\begin{equation}
\widetilde{\mathcal O}_{\Gamma}(z)
=
\overline{\psi}\left(z n_z\right)
\Gamma
W\left(z n_z,0\right)
\psi(0),
\qquad
n_z^\mu=(0,0,0,1),
\label{eq:equal_time_operator}
\end{equation}
where
\begin{equation}
W\left(zn_z,0\right)
=
\mathcal P
\exp\left[
-g\int_{0}^{z} d\lambda\,
A_z(\lambda n_z)
\right]
\end{equation}
is the straight spatial Wilson line connecting the two quark fields.
With suitable choices of $\Gamma$ and external states,
the Fourier transforms of forward hadron, vacuum-to-hadron,
and off-forward hadron matrix elements define quasi-PDFs,
quasi-DAs, and quasi-GPDs, respectively.
These quasi-distributions can then be perturbatively matched
onto their light-cone counterparts at large hadron momentum.

On the lattice, the renormalization of equal-time Wilson-line
operators involves several complications. The Wilson-line self-energy introduces a linear divergence,
$\delta m(a)\sim 1/a$, which exponentiates with the
Wilson-line length
~\cite{Brandt:1981kf,Chen:2016fxx}.
The nonlocal operator also contains logarithmic ultraviolet
divergences, which can be renormalized multiplicatively
~\cite{Ji:2017oey,Ishikawa:2017faj}.
The subtraction of the linear divergence also leaves a
finite mass ambiguity of order $\Lambda_{\rm QCD}$,
associated with the linear renormalon in perturbation
theory~\cite{Holligan:2023rex}. Its treatment must be consistent with the
prescription adopted for the perturbative matching
coefficients.
Furthermore, the reduced symmetries of the lattice
and explicit chiral-symmetry breaking can give rise to additional operator mixing,
including mixing between different Dirac structures
and with higher-dimensional operators
\cite{Constantinou:2017sej,Green:2017xeu,Chen:2017mie,Gockeler:1996mu}.
Matrix elements renormalized in an intermediate scheme can then be
related to $\MSbar$ through a finite scheme conversion.

Several renormalization schemes have been developed for
equal-time Wilson-line operators. The linear divergence
can be removed through a Wilson-line mass subtraction
~\cite{Chen:2016fxx}.
The regularization-independent momentum-subtraction
(RI/MOM) scheme determines the renormalization factors
from off-shell quark matrix elements, followed by a
perturbative conversion to the $\overline{\rm MS}$ scheme
~\cite{Martinelli:1994ty,Stewart:2017tvs,Alexandrou:2017huk}.
Alternatively, the ratio scheme cancels the multiplicative
ultraviolet divergences by dividing the matrix element
by its zero-momentum counterpart, with the conversion
to $\overline{\rm MS}$ based on the short-distance
operator product expansion (OPE)
~\cite{Radyushkin:2017cyf}.
At large Wilson-line separations, however, both RI/MOM
and the ratio scheme can suffer from nonperturbative
contamination~\cite{Ji:2020brr,LatticePartonLPC:2021gpi,Hua:2020gnw}.
In addition, RI/MOM requires precise gauge fixing,
particularly on fine lattices and for long Wilson lines
~\cite{Zhang:2024omt}.

The hybrid renormalization scheme was introduced to avoid
such long-distance contamination by treating the short-
and long-distance regions separately~\cite{Ji:2020brr}.
At short distances, the ratio scheme is
commonly used, where the short-distance OPE allows a
perturbative conversion to $\overline{\rm MS}$.
For the long-distance part, the hybrid scheme can be combined with
self-renormalization, which determines the divergent and finite
contributions by fitting the lattice-spacing and separation dependence of
bare zero-momentum matrix elements
\cite{LatticePartonLPC:2021gpi,LatticeParton:2022zqc}.
A commonly used parametrization is
\begin{align}
    \ln M(z,a)
    ={}&
    \frac{k|z|}
    {a\ln(a\Lambda_{\rm QCD})}
    +g(z)+f(z)a^p
    \nonumber\\
    &+
    \frac{3C_F}{\beta_0}
    \ln\left[
        \frac{\ln[1/(a\Lambda_{\rm QCD})]}
             {\ln[\mu/\Lambda_{\rm QCD}]}
    \right]
    +
    \ln\left[
        1+\frac{d}{\ln(a\Lambda_{\rm QCD})}
    \right],
    \label{eq:self_ren_fit}
\end{align}
where $\beta_0=11-2N_f/3$. The first term describes the linearly divergent
Wilson-line self-energy, including its leading
running-coupling dependence.
The function $g(z)$ contains the lattice-spacing-independent
finite contribution, including the residual mass term,
while $f(z)a^p$ parametrizes discretization effects.
The last two terms account for the leading and subleading
logarithmic dependence on the lattice spacing.
Although these contributions have distinct
lattice-spacing dependences, their separation requires
a simultaneous fit to data at multiple lattice spacings
and can be sensitive to the fit ansatz and discretization
effects. Furthermore, the finite residual mass $m_0$
must still be fixed by matching to the perturbative
$\overline{\rm MS}$ result at short distances.

\subsection{Gradient-flow ratio renormalization of quasi-distributions}
\label{subsec:flow-quasi-renormalization}

The separation of the Wilson-line self-energy, logarithmic renormalization,
and finite residual mass discussed in Sec.~\ref{subsec:quasi-renormalization}
also determines the structure of GFRR.
With the flowed fields defined in Sec.~\ref{subsec:gradient-flow}, the Wilson-line self-energy is regulated by the physical flow radius $\rF$
and its leading power-divergent contribution is determined
nonperturbatively from ratios at different flow radii.
The remaining conversion to $\MSbar$ requires perturbative matching and a
prescription for the finite residual mass.
We first formulate these relations at the operator level, separating the
Wilson-line mass term from the matching coefficients and their
renormalization-group evolution.
We then apply them to hadronic matrix elements normalized at $z=0$
and specify the subtraction, limiting procedure, and connection to the hybrid
scheme introduced in Sec.~\ref{subsec:quasi-renormalization}.

\subsubsection{Flowed operators and perturbative matching}
\label{subsubsec:flowed-operator-matching}

Using the ringed fields introduced in Sec.~\ref{subsec:gradient-flow},
we define the flowed quark bilocal by
\begin{equation}
  \widetilde{\mathcal O}^{\rm GF}_{\Gamma}(z,\tau)
  =\mathring{\bar\chi}(z n_z,\tau)\,
   \Gamma W_\tau(z n_z,0)\,
   \mathring{\chi}(0,\tau),
  \label{eq:gf-bilocal-operator}
\end{equation}
where the quark fields and the Wilson line are evaluated at the same flow
time. At fixed nonzero $z$, the SFTX takes the
form~\cite{Brambilla:2023vwm}
\begin{equation}
\widetilde{\mathcal O}^{\rm GF}_{\Gamma}(z,\tau)
  = e^{\delta m(\tau)|z|}C_\Gamma(z,\tau,\mu)\,
   \widetilde{\mathcal O}^{\MSbar}_{\Gamma}(z,\mu)+\order(\tau).
\label{eq:GF_MS_matching_full}
\end{equation}
After factoring out the Wilson-line mass term, the one-loop
matching coefficient reads~\cite{Monahan:2017hpu,Brambilla:2023vwm}
\begin{align}
C_\Gamma(z,\tau,\mu)  =1+&\frac{\alpha_s(\mu)C_F}{4\pi}
  \bigg[
    -3\ln\!\left(2\mu^2\tau e^{\gamma_E}\right)
    -2-\ln432
    \nonumber\\
    &-4e^{-\bar z^2}
    -3\operatorname{Ei}(-\bar z^2)
+4\sqrt{\pi}\,\bar z
    \bigl(1-\operatorname{erf}(\bar z)\bigr)
    +d_\Gamma(\bar z)
  \bigg]
  +\order(\alpha_s^2),
  \label{eq:gf-finite-flow-matching}
\end{align}
where
\begin{equation}
  d_\Gamma(\bar z)=
  \begin{cases}
    \displaystyle
    \frac{3(1-e^{-\bar z^2})}{\bar z^4}
    -\frac{2+e^{-\bar z^2}}{\bar z^2},
    & \Gamma=\gamma_z,\\[8pt]
    \displaystyle
    -\frac{1-e^{-\bar z^2}}{\bar z^4}
    +\frac{2-e^{-\bar z^2}}{\bar z^2},
    & \Gamma=\gamma_\perp^\mu.
  \end{cases}
  \label{eq:gf-dirac-dependent-correction}
\end{equation}
Here, $\bar z=|z|/\rF$, and $\operatorname{Ei}(x)$ and
$\operatorname{erf}(x)$ denote the exponential integral
and the error function, respectively.

For fixed nonzero $z$, the $\bar z$-dependent terms in
Eq.~\eqref{eq:gf-finite-flow-matching} vanish as $\rF/|z|\to0$.
The matching coefficient then reduces to
\begin{equation}
  C_q(\tau,\mu)
  =1-\frac{\alpha_s(\mu)C_F}{4\pi}
    \left[
      3\ln\!\left(2\mu^2\tau e^{\gamma_E}\right)
      +2+\ln432
    \right]
    +\order(\alpha_s^2),
  \label{eq:Cq_one_loop}
\end{equation}
which no longer depends on $z$. The leading SFTX can therefore be
written as~\cite{Brambilla:2023vwm}
\begin{equation}
\widetilde{\mathcal O}^{\rm GF}_{\Gamma}(z,\tau)
  =  e^{\delta m(\tau)|z|}
  C_q(\tau,\mu)\,
   \widetilde{\mathcal O}^{\MSbar}_{\Gamma}(z,\mu)
   +\order(\tau),
  \label{eq:GF_MS_matching}
\end{equation}
with
\begin{equation}
  \delta m(\tau)
  =-\frac{\alpha_s(\mu)C_F}{4\pi}
    \frac{\sqrt{2\pi}}{\sqrt{\tau}}
    +\order\!\left(\frac{\alpha_s^2}{\sqrt{\tau}}\right).
  \label{eq:gf-line-mass}
\end{equation}

In lattice calculations of quasi-distributions, the nonlocal
matrix elements are often normalized by their corresponding
local-current matrix elements. Besides reducing common lattice
artifacts, this normalization cancels the 
wave-function renormalization. In particular, for gradient-flow calculations, this normalization avoids the explicit determination of the ringed-field normalization factor defined in Eq.~\eqref{eq:ringed_Z}, whose direct lattice evaluation involves disconnected fermion loops and can suffer from large statistical uncertainties.

For the matrix elements normalized at $z=0$,
the SFTX also requires the matching coefficient of the corresponding local
current. Defining
$J_\Gamma^{\rm GF}(\tau)=
\widetilde{\mathcal O}^{\rm GF}_\Gamma(0,\tau)$
and
$J_\Gamma^{\MSbar}(\mu)=
[\bar\psi(0)\Gamma\psi(0)]_{\MSbar}$,
we have
\begin{align}
  J_\Gamma^{\rm GF}(\tau)
  &=C_{\rm loc}(\tau,\mu)
    J_\Gamma^{\MSbar}(\mu)+\order(\tau),
  \label{eq:gf-local-current-matching}\\
  C_{\rm loc}(\tau,\mu)
  &=1+\frac{\alpha_s(\mu)C_F}{4\pi}
    \left(\frac12-\ln432\right)
    +\order(\alpha_s^2),
  \label{eq:gf-local-current-coefficient}
\end{align}
for the vector and flavor-nonsinglet axial currents
~\cite{Hieda:2016lly}. Combining this with the nonlocal
matching coefficient gives $C_{\rm R}=C_q/C_{\rm loc}$,
\begin{align}
  C_{\rm R}(\tau,\mu)
  =1-\frac{\alpha_s(\mu)C_F}{4\pi}
    \left[
      3\ln\!\left(2\mu^2\tau e^{\gamma_E}\right)
      +\frac52
    \right]+\order(\alpha_s^2).
  \label{eq:gf-normalized-matching}
\end{align}
Through standard renormalization group (RG) evolution, one arrives at the
RG-resummed matching kernel
\begin{equation}
  C_{\rm R}^{\rm RG}(\tau,\mu)
  =
  \left[
    \frac{\alpha_s(\mu)}
         {\alpha_s(\mu_\tau)}
  \right]^{3C_F/\beta_0}
   C_{\rm R}(\tau,\mu_\tau),
  \label{eq:gf-rg-matching}
\end{equation}
where the flow scale $\mu_\tau=e^{-\gamma_E/2}/\sqrt{2\tau}$ is chosen to eliminate the logarithm in the one-loop matching coefficient. 

Besides logarithmic matching, the linear divergence
associated with the Wilson-line self-energy must also
be addressed. Although $\delta m(\tau)$ is perturbatively
calculable, the perturbative expansion of the static
self-energy in conventional lattice regularization
is known to exhibit asymptotic behavior associated
with renormalons
\cite{Pineda:2001zq,Bauer:2011ws,Bali:2013pla}.
Similar convergence issues also persist in the
gradient-flow scheme, motivating a nonperturbative
determination of $\delta m(\tau)$~\cite{Zhang:2025mer}.  A practical advantage over the conventional lattice regulator is that
the Wilson-line mass correction can be studied continuously as a
function of the flow time. In the usual lattice regularization, the coefficient
of the $1/a$ divergence depends on details such as the lattice action
and smearing prescription, and its determination normally requires
several lattice spacings~\cite{LatticePartonLPC:2021gpi}.  With
gradient flow, one can instead use the $\tau$ dependence of
$\delta m(\tau)$ on each ensemble.  A continuum
extrapolation at fixed physical flow time is still required, but the
result is then defined in the common gradient-flow scheme rather than by
a particular lattice regulator.  In this sense, the linear divergence
in gradient flow is a more systematically controlled quantity than the conventional
$1/a$ divergence.

These operator-level relations determine the renormalization of flowed
hadronic matrix elements. The following subsection formulates the
corresponding prescription for the quasi-DA and quasi-PDF channels
considered in this work.

\subsubsection{Renormalization of hadronic matrix elements}
\label{subsubsec:flowed-hadronic-renormalization}

In this work, we study the pion quasi-DA and the unpolarized pion and proton
quasi-PDFs. Using unringed flowed quark fields, the lattice matrix
elements are
\begin{align}
  H_\pi^{\rm DA}(z,P_z,\rF,a)
  &=\langle0|\bar\chi_d(z n_z,\tau)\gamma_t\gamma_5
    W_\tau(z n_z,0)\chi_u(0,\tau)|\pi(P_z)\rangle,
  \label{eq:pion-da-flowed}\\
  H_{\pi}^{\rm PDF}(z,P_z,\rF,a)
  &=\langle\pi(P_z)|\bar\chi_q(z n_z,\tau)\gamma_t
    W_\tau(z n_z,0)\chi_q(0,\tau)|\pi(P_z)\rangle,
  \label{eq:pion-pdf-flowed}\\
  H_{p}^{\rm PDF}(z,P_z,\rF,a)
  &=\langle p(P_z)|\bar\chi_q(z n_z,\tau)\gamma_t
    W_\tau(z n_z,0)\chi_q(0,\tau)|p(P_z)\rangle,
  \label{eq:proton-pdf-flowed}
\end{align}
where we focus on the isovector combination $u-d$ for quasi-PDF.
For the pion quasi-DA, we choose $\gamma_t\gamma_5$ to project the light-cone distribution at large momentum, which is a commonly used choice in previous
calculations~\cite{LatticeParton:2023xdl,LatticePartonLPC:2023pdv}. 

For Wilson-type fermions, these operators for quasi-PDF and quasi-DA exhibit finite mixing due to explicit chiral symmetry breaking at nonzero lattice spacing~\cite{Constantinou:2017sej}. Such mixing is expected to vanish in the continuum limit at fixed physical flow radius~\cite{Kim:2021qae}.

We normalize each matrix element by its value at $z=0$,
\begin{align}
  \Mcal_\pi^{\rm DA}(z,P_z,\rF,a)
  &=\frac{H_\pi^{\rm DA}(z,P_z,\rF,a)}
          {H_\pi^{\rm DA}(0,P_z,\rF,a)},
  \label{eq:da-local-normalization}\\
  \Mcal_h^{\rm PDF}(z,P_z,\rF,a)
  &=\frac{H_h^{\rm PDF}(z,P_z,\rF,a)}
          {H_h^{\rm PDF}(0,P_z,\rF,a)},
  \qquad h\in\{\pi,p\}.
  \label{eq:local-normalization}
\end{align}
For the pion DA,
this removes the factor of $f_\pi$ associated
with the axial current, while for the isovector
PDF, the normalization is fixed by the conserved
vector charge.

Denoting the
three channels collectively by $X$, their leading flow dependence
can be written schematically as
\begin{align}
  \Mcal_X(z,P_z,\rF,a)
  ={}&C_{\rm R}(\tau,\mu)
   \exp\!\left[-\frac{A_F(\rF)}{\rF}|z|-m_{0,F}|z|\right]
   \Mcal_X^{\MSbar}(z,P_z,\mu)\nonumber\\
   &+\order\!\left(\frac{\tau}{z^2},\tau P_z^2,
                   \tau\Lambda_{\rm QCD}^2\right),
  \label{eq:schematic-factorization}
\end{align}
where $A_F$ parametrizes the leading $1/\rF$
divergence of the Wilson-line self-energy,
while $m_{0,F}$ denotes the finite residual mass.
Although $A_F$ may depend logarithmically on $\rF$,
we approximate it by an effective constant over
the flow-radius range used in our numerical fits.

The state independence of the Wilson-line self-energy
allows it to be extracted from various observables,
including the hadronic matrix elements considered above,
straight Wilson lines, Wilson loops, and Polyakov-loop
correlators. For the hadronic matrix elements, we seek
a cleaner determination of its flow-radius dependence
by constructing a double ratio at two flow radii.
This cancels the physical matrix element and the
flow-independent residual mass, while the remaining
matching factors are independent of $z$ at leading
power and contribute only to the overall normalization.
The resulting $|z|$ dependence thus isolates the
difference of the Wilson-line masses at the two
flow radii, as discussed in
Sec.~\ref{sec:self-energy}.

With the linear divergence subtracted, the matrix
elements can be extrapolated to the continuum limit.
The remaining flow-time dependence can then be
removed through perturbative matching, followed
by the small-flow-time limit. At fixed nonzero separation, the power-subtraction and matching steps
of GFRR can be expressed as
\begin{align}
  \Mcal_X^{(0)}(z,P_z,\mu)
  &\equiv\lim_{\tau\to0}\left[
    \frac{1}{C_{\rm R}^{\rm RG}(\tau,\mu)}
    \lim_{a\to0}\left\{
      e^{A_F(\rF)|z|/\rF}
      \frac{H_X(z,P_z,\rF,a)}{H_X(0,P_z,\rF,a)}
    \right\}\right]
    \nonumber\\
  &=e^{-m_{0,F}|z|}\Mcal_X^{\MSbar}(z,P_z,\mu).
  \label{eq:framework-renormalization-map}
\end{align}
The last equality holds to the matching accuracy used here.
The limits require a window with $a\ll\rF\ll|z|$ and small flow
corrections from external physical scales. With the two ensembles
available here, we compare cutoff dependence at fixed $\rF$;
a controlled continuum extrapolation remains to be performed. The residual mass $m_{0,F}$ is then  determined by fitting the
zero-momentum matrix element to its short-distance OPE
with the leading-renormalon-resummed (LRR) Wilson
coefficient~\cite{Holligan:2023rex}.

Finally, the hybrid construction joins
short-distance ratio renormalization to the mass-subtracted
long-distance result~\cite{Ji:2020brr}.
Sec.~\ref{sec:step-by-step} presents the numerical implementation.
The resulting renormalized coordinate-space matrix elements can subsequently
be Fourier transformed and matched to light-cone distributions.

\section{Nonperturbative Wilson-line self energy}
\label{sec:self-energy}

The Wilson-line self-energy gives rise to a linear divergence
proportional to $1/a$ in the absence of gradient flow.
With gradient flow, the corresponding factor in the continuum
takes the form $\exp[-A_R|z|/\rF]$ at small flow time
\cite{Luscher:2011bx,Monahan:2016bvm,Monahan:2017hpu,
Brambilla:2023vwm}.
The coefficient $A_R$ depends on the color representation $R$
but is independent of the external hadron and its momentum.
At fixed physical $\rF$, its lattice-spacing dependence
arises only from discretization effects.
As discussed in Sec.~\ref{subsubsec:flowed-hadronic-renormalization},
$A_R$ may depend logarithmically on $\rF$.
In the following, we treat $A_R$ as a constant over the
flow-radius range used in the fits and examine its universality
using different observables.
The resulting coefficient supplies the nonperturbative power subtraction
in GFRR.

The independent variation of the lattice spacing $a$ and the flow radius $\rF$
offers two complementary ways to probe this ultraviolet structure. Comparing
different ensembles at the same physical $\rF$ tests whether the leading
exponential falloff is governed by the physical flow scale rather than the
lattice cutoff $a$. Conversely, varying $\rF$ on a single ensemble changes the
ultraviolet smoothing scale on the exact same gauge configurations, allowing us
to extract $A_R$ without altering the long-distance physics. We carry out this
study on two $N_f=2+1$ tadpole-improved clover ensembles generated by the CLQCD
collaboration \cite{CLQCD:2023sdb}, C24P29 and F32P30, summarized in
Table~\ref{tab:lattice-setup}. On both ensembles, we evaluate rectangular
Wilson loops, Landau-gauge straight Wilson lines, and hadronic quasi matrix
elements across the same sequence of physical flow radii,
$\rF=0, 0.06, 0.08, \ldots, 0.20\,\mathrm{fm}$. 
Statistical correlations between different flow times
are taken into account in the jackknife analysis.

\begin{table}[t]
  \centering
  \caption{Ensembles and numerical ranges used in the self-energy analysis.
  The configuration counts are the complete extended samples used for the
  quasi and straight-Wilson-line measurements across the common sequence of
  physical flow radii $\rF$. The lattice size is $N_s^3\times N_t$,
  with physical spatial extent $L=N_s a$.}
  \label{tab:lattice-setup}
  \begin{tabular}{lcc}
    \toprule
      & C24P29 & F32P30 \\
    \midrule
    $a$ [fm] & $0.10530(18)$ & $0.07746(18)$ \\
    $N_s^3\times N_t$ & $24^3\times72$ & $32^3\times96$ \\
    $m_\pi$ [MeV] & $292.4(1.1)$ & $303.2(1.3)$ \\
    configurations & 961 & 886 \\
    $\rF$ [fm] & \multicolumn{2}{c}{$0,0.06,0.08,\ldots,0.20$} \\
    quasi separation & $|z|/a=0,\ldots,11$ & $|z|/a=0,\ldots,16$ \\
    quasi $P_z$ [GeV]
      & $0,0.491,\ldots,2.453$ & $0,0.500,\ldots,2.501$ \\
    \bottomrule
  \end{tabular}
\end{table}
\FloatBarrier

\subsection{Flow-radius scaling in quasi matrix elements}
\label{subsec:quasi-scale-transfer}

In a hadronic quasi matrix element, the spatial falloff with $|z|$ combines the
nonperturbative long-distance spatial dependence of the hadronic matrix element
with the ultraviolet self-energy suppression of the Wilson line. Before
separating these two pieces quantitatively, we can already observe the transfer
of the ultraviolet scale from $a$ to $\rF$ directly in the matrix elements. We
evaluate the pion quasi-DA, the pion quasi-PDF, and the unpolarized proton
quasi-PDF at momentum modes $P_zL/(2\pi)=0,\ldots,5$ on two ensembles.

For the hadronic matrix elements used in the self-energy analysis, we combine
temporal fits using Akaike information criterion (AIC) weights.
For the pion quasi-DA, the two-state extractions use temporal lower bounds
$t_{\min}/a=9,10$.
For the pion and proton quasi-PDFs, we retain source--sink separations
$t_{\rm seq}/a\ge9$ and average over symmetric temporal cuts
$t_{\rm cut}/a=1,2,3$ near the source and sink.

\begin{figure}[t]
  \centering
  \includegraphics[width=0.98\textwidth]{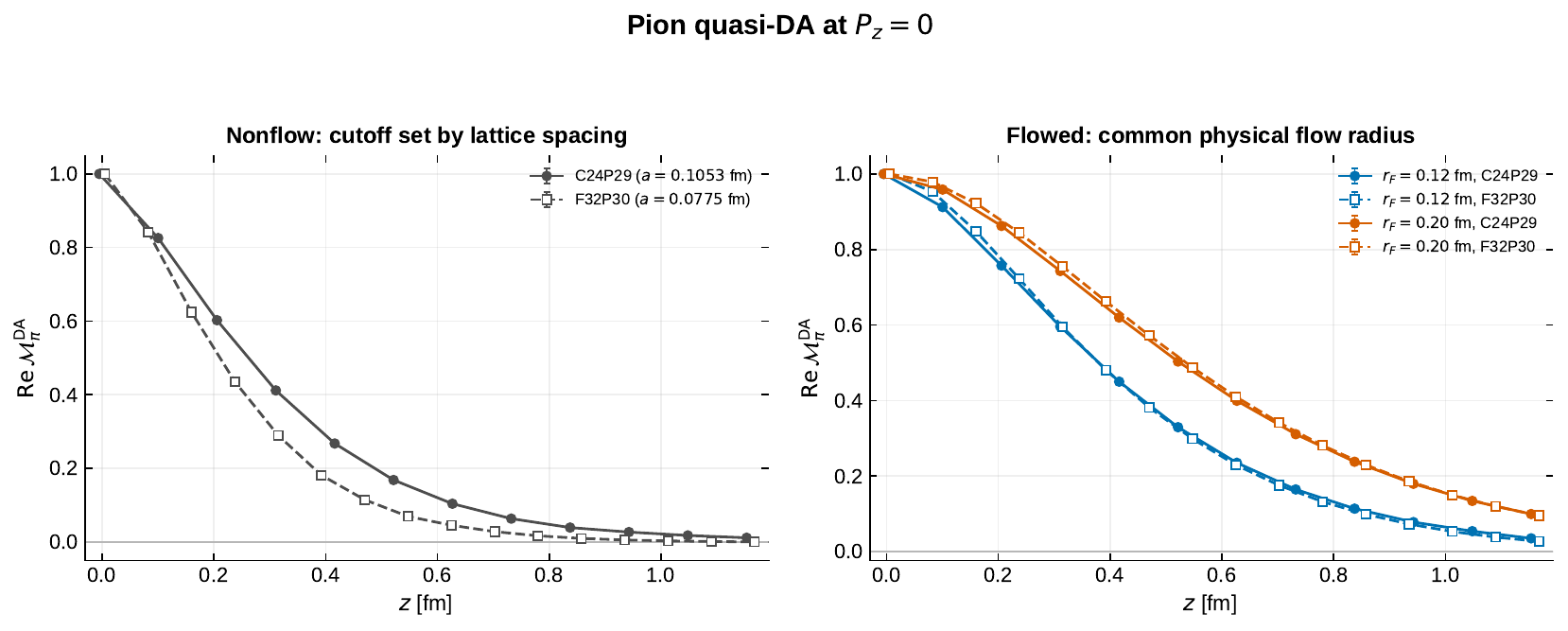}
  \caption{Real part of the $P_z=0$ pion quasi-DA matrix element normalized at
  $z=0$. Left: nonflowed ($\tau=0$) results on the two ensembles, where the
  ultraviolet decay is controlled by the lattice spacing $a$. Right: flowed
  results at common physical radii $\rF=0.12$ and $0.20\,\mathrm{fm}$. Matching
  the physical flow radius strongly reduces the separation between the two
  lattice spacings, while varying $\rF$ moves both ensembles coherently.}
  \label{fig:uv-scale-transfer}
\end{figure}

Fig.~\ref{fig:uv-scale-transfer} illustrates this scale transfer using the
real part of the bare $P_z=0$ pion quasi-DA matrix element normalized at $z=0$,
prior to any self-energy subtraction. In the unflowed case ($\tau=0$,
left panel), the Wilson-line divergence is regulated by the lattice spacing $a$
and scales as $\sim\exp[-c|z|/a]$. As a result, the C24P29 and F32P30 curves
split apart immediately as $z$ increases, with the finer ensemble F32P30
($a=0.0775\,\mathrm{fm}$) decaying noticeably faster than the coarser ensemble
C24P29 ($a=0.1053\,\mathrm{fm}$). Once the fields are evolved to a common
physical flow radius (right panel), this cutoff dependence largely disappears:
at both $\rF=0.12\,\mathrm{fm}$ (blue) and $\rF=0.20\,\mathrm{fm}$ (orange),
the results from the two ensembles nearly overlap across the entire spatial
range up to $z\approx 1.15\,\mathrm{fm}$, leaving small residual ensemble
differences at short distances consistent with suppressed $\order(a^2/\rF^2)$
discretization effects. Meanwhile, reducing the physical smoothing radius from
$\rF=0.20\,\mathrm{fm}$ to $0.12\,\mathrm{fm}$ shifts both ensembles downward
together and steepens their spatial falloff, similar to the effect of reducing
$a$.

\FloatBarrier

This comparison confirms directly in the hadronic matrix element that gradient
flow replaces $a$ with $\rF$ as the dominant ultraviolet scale. However, the
spatial decay at a fixed flow radius still entangles the $\exp[-A_F|z|/\rF]$
factor with the unknown spatial dependence of the hadronic matrix element, so
the two cannot be separated from a single-radius profile alone without modeling
the long-distance physics. We therefore turn first to pure-gauge observables to
determine the Wilson-line self energy independently, before using flow double
ratios to extract and cross-check it in the hadronic channels.

\subsection{Gauge-invariant determination from the static potential}
\label{subsec:static-potential}

To determine the Wilson-line self energy quantitatively, a standard
gauge-invariant route is through the static quark-antiquark potential
\cite{Chen:2016fxx,Brambilla:2023fsi}. Forming a closed gauge contour removes
the endpoint gauge transformations that make the expectation value of an open
line vanish without gauge fixing, while the Euclidean time evolution projects
out the ground-state energy of a static quark-antiquark pair along with the
self energies of the temporal Wilson lines. On the lattice, this
gauge-invariant static potential can be extracted either from the large-$T$
exponential decay of closed rectangular Wilson loops, or from spatial
correlators of temporal Polyakov loops.

\begin{figure}[t]
  \centering
  \includegraphics[width=0.99\textwidth]{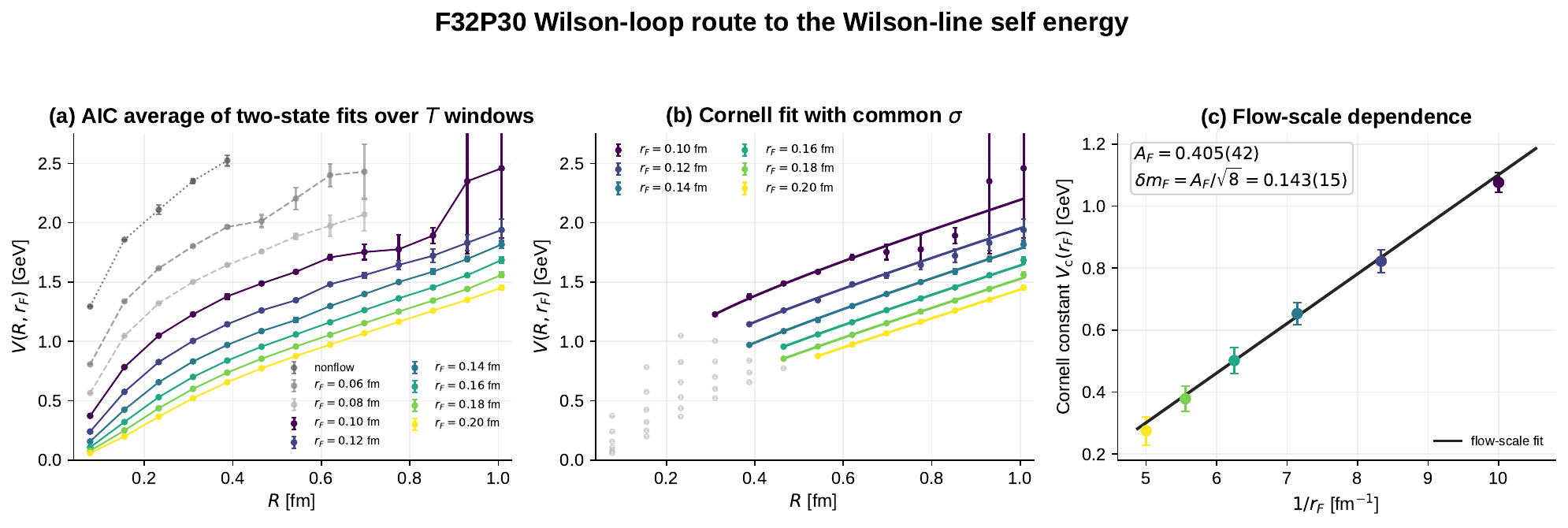}
  \caption{Gauge-invariant static-potential determination on F32P30. (a) Static
  potentials obtained from an AIC average of two-state Wilson-loop fits across
  temporal windows, with the nonflow result and those at the two smallest flow
  radii shown in gray for comparison. (b) Joint Cornell fit with a common string tension $\sigma$
  over the flow-radius-dependent spatial window $2\rF+a<R\le R_{\max}$
  ($R_{\max}\approx 1.0\,\mathrm{fm}$); faint gray points lie outside the fit
  window and are excluded. (c) Additive constants $V_{\rm c}(\rF)$ versus
  $1/\rF$, whose slope contains two Wilson-line self energies. Dividing by two
  and converting with $\delta m_F=A_F/\sqrt8$ gives the single-line value
  quoted in the text.}
  \label{fig:static-potential}
\end{figure}

In the rectangular Wilson-loop approach, a loop $W(R,T;\rF)$ of spatial width
$R$ and temporal extent $T$ constructed from the flowed gauge field is fitted
using the truncated spectral decomposition
\begin{equation}
  \langle W(R,T;\rF)\rangle
  =\sum_{n=0}^{N_{\rm exp}-1}Z_n(R,\rF)
   e^{-V_n(R,\rF)T},
  \label{eq:wilson-loop-potential}
\end{equation}
where $N_{\rm exp}$ is the number of exponential terms retained in the fit
($N_{\rm exp}=2$ in our analysis), and the ground-state energy $V_0(R,\rF)$
defines the static potential at flow radius $\rF$. To suppress sensitivity to
the lower bound $T_{\min}$ of the temporal fit window when removing
excited-state contamination, we perform two-state fits across several temporal
windows and combine the results using AIC weights. The resulting static potentials are shown in
Fig.~\ref{fig:static-potential}(a).

At intermediate spatial separations $R$, the resulting static potential is
described by the Cornell parametrization,
\begin{equation}
  V(R,\rF)=\sigma R-\frac{e(\rF)}{R}+V_{\rm c}(\rF).
  \label{eq:cornell-potential}
\end{equation}
On the short-distance side, the separation $R$ must be sufficiently larger than
the flow radius to reduce short-distance distortions of the $R$ dependence
induced by gradient flow~\cite{Brambilla:2023fsi}; we therefore impose the
flow-radius-dependent lower bound $R>2\rF+a$. This condition reduces flow
distortions in the $R$ dependence of the potential, whereas the flow-radius
dependence in the additive constant $V_{\rm c}(\rF)$ remains and enters the
self-energy extraction below. On the long-distance side, string breaking in
dynamical QCD~\cite{Bulava:2019iut} restricts the range over which the linear
Cornell term applies. On F32P30, we restrict the spatial fit window to
$2\rF+a<R\le R_{\max}$ with $R_{\max}\approx 1.0\,\mathrm{fm}$, and adopt a
joint Cornell parametrization across the flow-radius range
$\rF\in[0.10,0.20]\,\mathrm{fm}$ with a common string tension $\sigma$, while
allowing the Coulomb coefficient $e(\rF)$ and additive constant $V_{\rm c}(\rF)$
to vary with $\rF$, as shown in Fig.~\ref{fig:static-potential}(b). The
additive constant contains the ultraviolet self energies of the two temporal
Wilson lines of length $T$ in the rectangular loop, so its leading-order
flow-radius dependence takes the form
\begin{equation}
  V_{\rm c}(\rF)=2A_F\frac{\hbar c}{\rF}+c_0,
  \qquad \delta m_F\equiv\frac{A_F}{\sqrt8}>0,
  \label{eq:cornell-flow}
\end{equation}
where the factor of two accounts for the two temporal Wilson lines in the loop.
Fitting $V_{\rm c}(\rF)$ linearly in $1/\rF$ on F32P30, as shown in
Fig.~\ref{fig:static-potential}(c), gives
\begin{equation}
  A_F^{V}=0.405(42),
  \qquad
  \delta m_F^{V}=0.143(15).
  \label{eq:static-potential-result}
\end{equation}
Built entirely from closed Wilson loops, this determination is manifestly gauge
invariant.

Similarly, the static potential and its additive self-energy constant can also
be accessed from spatial correlators of flowed Polyakov loops,
\begin{equation}
  P(\bm x;\rF)=\frac{1}{N_c}\operatorname{Tr}
  \prod_{t=0}^{N_t-1}U_t(\bm x,t;\rF),
  \qquad
  C_P(R;\rF)=\big\langle P(\bm x;\rF)
  P^\dagger(\bm x+\bm R;\rF)\big\rangle
  \propto e^{-V(R,\rF)\beta},
  \label{eq:polyakov-potential}
\end{equation}
with $\beta=aN_t$ denoting the physical temporal extent of the lattice and
$N_t$ the number of temporal lattice sites (Table~\ref{tab:lattice-setup}).
Because a Polyakov loop wraps around the entire periodic temporal extent, this
construction does not require choosing a temporal fit window as in the
rectangular Wilson-loop analysis. Extracting the zero-temperature ground-state
potential from Eq.~\eqref{eq:polyakov-potential} still requires a sufficiently
large temporal extent $\beta$ with ground-state dominance, whereas at finite
temperature the correlator determines the static quark-antiquark free energy,
which carries the same ultraviolet Wilson-line self
energy~\cite{Brambilla:2010xn}. Polyakov-loop correlators are therefore
particularly useful in finite-temperature studies with shorter physical
temporal extents or in quenched calculations where high statistics can be
accumulated. For the zero-temperature $N_f=2+1$ ensembles used here ($N_t\times a=7.58$
and $7.44$ fm), we employ the rectangular Wilson-loop analysis above to determine
the static potential.

\subsection{Landau-gauge straight Wilson lines}
\label{subsec:gauge-fixed-line}

In the static-potential analysis of Sec.~\ref{subsec:static-potential}, the
self energy is extracted through a sequence of fits—spectral decomposition in
$T$, a Cornell fit in $R$, and a linear fit in $1/\rF$. Alongside this
gauge-invariant route, we can also examine a more direct extraction by measuring
the open straight spatial Wilson line after gauge fixing. Although the overall
expectation value of an open Wilson line depends on the gauge choice, its
leading length-dependent ultraviolet suppression $\exp[-A_R|z|/\rF]$ arises
from short-distance fluctuations along the contour and shares the same
coefficient $A_R$ as in a gauge-invariant operator.

We choose the Landau gauge in this work, which preserves Euclidean rotational
symmetry and is the standard covariant choice for nonperturbative nonlocal
renormalization studies. For each flowed gauge field, we determine the gauge
transformation $g(x)$, acting as
\begin{equation}
  U_\mu^g(x,\tau)
  =g(x)U_\mu(x,\tau)g^\dagger(x+\hat\mu),
\end{equation}
that maximizes the functional
\begin{equation}
  F[g;U(\tau)]
  =\frac{1}{4N_cV}\sum_{x,\mu}
    \operatorname{ReTr}U_\mu^g(x,\tau).
  \label{eq:landau-functional}
\end{equation}
Convergence to the stationarity condition is monitored using
\begin{align}
  M(x)&=\sum_\mu\left[
    U_\mu^g(x)+U_\mu^{g\dagger}(x-\hat\mu)
    \right],\nonumber\\
  \Delta(x)&=M(x)-M^\dagger(x)
    -\frac{1}{N_c}\operatorname{Tr}\!\left[
      M(x)-M^\dagger(x)\right]\mathbf{1},\nonumber\\
  \theta&=\frac{1}{N_cV}\sum_x
    \operatorname{Tr}\!\left[\Delta(x)\Delta^\dagger(x)\right].
  \label{eq:landau-residual}
\end{align}
This definition follows the normalization used in QUDA.
Long Wilson lines on
fine lattices are known to be sensitive to incomplete gauge
fixing~\cite{Zhang:2024omt}; we therefore iterate our over-relaxation algorithm
to a strict tolerance of $\theta<10^{-14}$. Tightening this criterion further to
$10^{-15}$ changes the longest lines at $\rF=0.12$ and $0.20\,\mathrm{fm}$ by
less than $4\times10^{-7}$ on both C24P29 and F32P30, confirming that residual
gauge-fixing uncertainties are negligible compared with the statistical and
fit-window uncertainties below.

After gauge fixing, we measure the cubic average of straight spatial Wilson
lines in color representation $R$,
\begin{equation}
  L_R(z,\rF)
  =\frac{1}{3d_R V}\sum_{i=1}^{3}\sum_x
    \operatorname{ReTr}_R\!\left[
      \prod_{n=0}^{z/a-1}U_i^{(R)}(x+n\hat i,\tau)
    \right],
  \label{eq:gauge-fixed-line}
\end{equation}
where $d_R$ is the dimension of representation $R$ and $L_R(0,\rF)=1$ by
construction. Averaging over all lattice origins and the three spatial
orientations makes $L_R(z,\rF)$ a probe with high statistical precision.

\begin{figure}[t]
  \centering
  \includegraphics[width=0.99\textwidth]{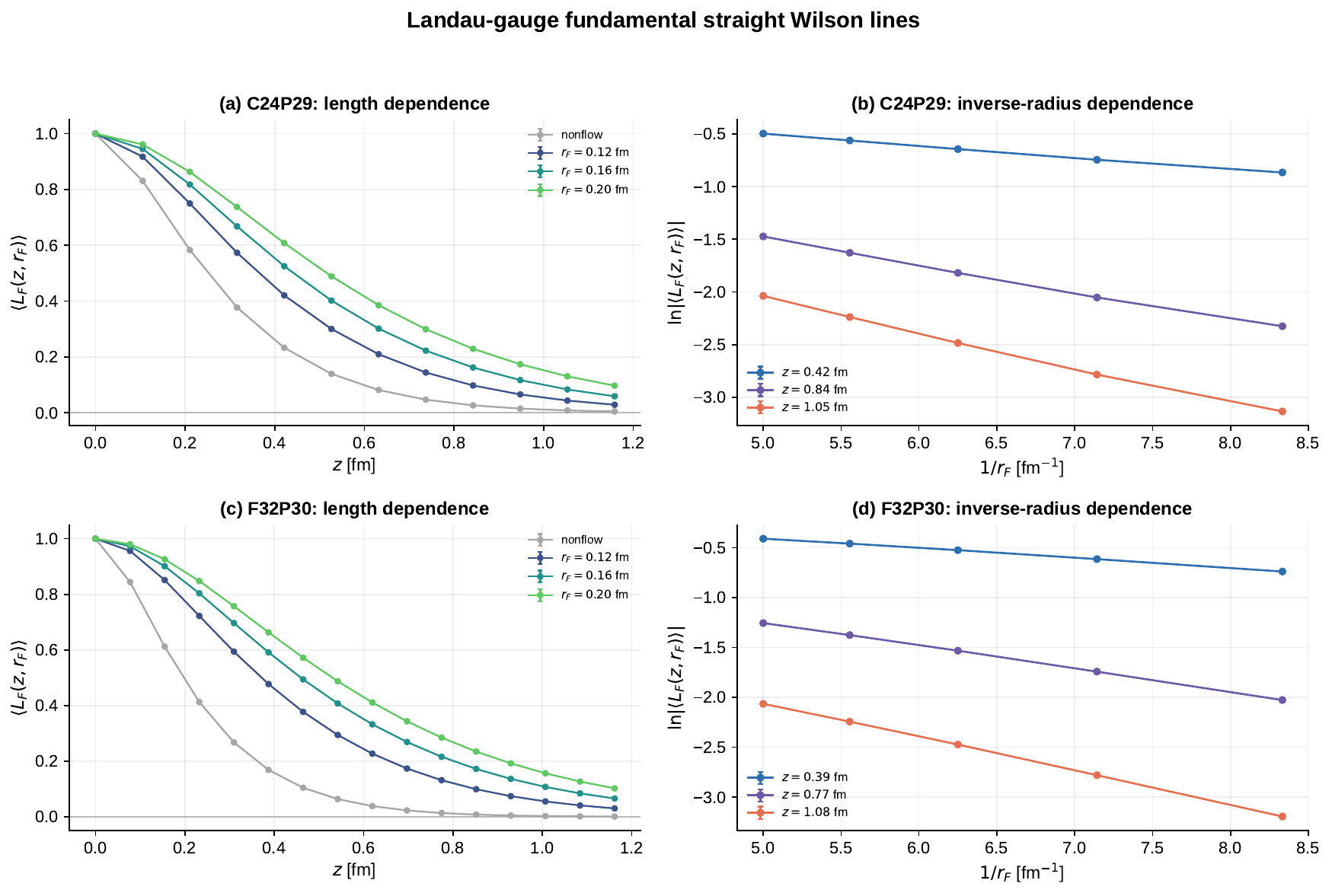}
  \caption{Cubic-averaged Landau-gauge straight Wilson lines on C24P29 (upper
  row) and F32P30 (lower row). Left: length dependence at representative flow
  radii. Right: $\ln|L_F|$ versus $1/\rF$ at representative physical lengths
  across $\rF=0.12$--$0.20\,\mathrm{fm}$, prior to forming any flow ratio.}
  \label{fig:raw-straight-line}
\end{figure}

Fig.~\ref{fig:raw-straight-line} displays the expectation values $L_F(z,\rF)$
on C24P29 and F32P30 prior to forming flow ratios. Both aspects of the expected
leading term $\exp[-A_F|z|/\rF]$ are clearly visible: at fixed $\rF$,
$L_F(z,\rF)$ decays approximately exponentially with $z$ (left panels), while
at fixed physical length $z$, $\ln|L_F(z,\rF)|$ is nearly linear in $1/\rF$
across $\rF=0.12$--$0.20\,\mathrm{fm}$ (right panels). However, fitting the $z$
dependence at a single $\rF$ still mixes the ultraviolet self energy with the
flow-independent long-distance behavior of the gauge-fixed line. To remove that
long-distance piece without modeling it, we next compare two flow radii at the
same spatial separation.

\subsection{Flow double ratio}
\label{subsec:double-ratio}

The straight Wilson-line data combine the ultraviolet exponential
$\exp[-A_R|z|/\rF]$ with endpoint factors and a flow-independent long-distance
function of $z$, while a hadronic quasi matrix element combines the same
exponential with a physical hadronic matrix element. In the regime
$a\ll\rF\ll|z|$, these long-distance $z$-dependent functions are insensitive to
the ultraviolet flow scale and cancel directly when the same observable is
compared at two different flow radii on the same configurations.

Specifically, for a straight Wilson line in representation $R$, the expectation
value factorizes as
\begin{equation}
  L_R(z,\rF)
  =Z_L(\rF)\exp\!\left[-A_R\frac{|z|}{\rF}\right]
   \ell_R(z)\,[1+\epsilon_L(z,\rF)],
  \label{eq:straight-line-factorization}
\end{equation}
where $\ell_R(z)$ contains the flow-independent long-distance behavior.
Similarly, following Eq.~\eqref{eq:schematic-factorization}, a normalized quasi
matrix element takes the factorized form
\begin{equation}
  \Mcal_X(z,P_z,\rF)
  =C_{\rm R}(\tau,\mu)\exp\!\left[-A_F\frac{|z|}{\rF}-m_0|z|\right]
   \Mcal_X^{\MSbar}(z,P_z,\mu)\,[1+\epsilon_X(z,\rF)],
  \label{eq:quasi-flow-factorization}
\end{equation}
where $\Mcal_X^{\MSbar}(z,P_z,\mu)$ is the renormalized matrix element in the $\MSbar$ scheme,
$m_0$ denotes the flow-independent residual mass $m_{0,F}$ of
Eq.~\eqref{eq:schematic-factorization}, and $\epsilon_{L,X}(z,\rF)$ denote
finite-flow and discretization corrections. We choose $\rref=0.20\,\mathrm{fm}$
as the reference flow radius—being the largest flow radius in our data set, it
provides the strongest statistical signal—and form the ratios on each jackknife
sample:
\begin{align}
  \Rcal_L(z;\rF,\rref)
  &=\frac{L_R(z,\rF)}{L_R(z,\rref)},
  \label{eq:line-double-ratio}\\
  \Rcal_X(z,P_z;\rF,\rref)
  &=\frac{\Mcal_X(z,P_z,\rF)}{\Mcal_X(z,P_z,\rref)}
   =\frac{H_X(z,P_z,\rF)/H_X(0,P_z,\rF)}
          {H_X(z,P_z,\rref)/H_X(0,P_z,\rref)}.
  \label{eq:quasi-double-ratio}
\end{align}
In Eq.~\eqref{eq:quasi-double-ratio}, the bare quasi matrix elements are first
normalized by their local values at $z=0$, and the flow double ratio $\Rcal_X$
then compares these normalized matrix elements at two different flow radii. In
both $\Rcal_L$ and $\Rcal_X$, the unknown long-distance functions $\ell_R(z)$
and $\Mcal_X^{\MSbar}(z,P_z,\mu)$ cancel without any modeling, and in $\Rcal_X$
the flow-independent residual mass $m_0$ cancels as well.

Taking the logarithm at each flow radius $\rF$, excluding the reference radius
$\rref$, gives a linear function of $|z|$ at leading order:
\begin{equation}
  \ln|\Rcal_{L,X}(z;\rF,\rref)|
  = B_{L,X}(\rF,\rref)
    -\Delta_m(\rF,\rref)|z|
    +\Delta\epsilon_{L,X}(z;\rF,\rref),
  \label{eq:double-ratio-linear-z}
\end{equation}
where
$\Delta\epsilon_{L,X}
=\ln\!\left|[1+\epsilon_{L,X}(z,\rF)]/
                 [1+\epsilon_{L,X}(z,\rref)]\right|$
collects the finite-flow and discretization corrections in the ratio.
The slope is determined by the self-energy difference between $\rF$ and $\rref$,
\begin{equation}
  \Delta_m(\rF,\rref)
  =A_R\left(\frac{1}{\rF}-\frac{1}{\rref}\right),
  \qquad
  \delta m_R\equiv\frac{A_R}{\sqrt{8}}>0.
  \label{eq:flow-law}
\end{equation}
Here the free intercept $B_{L,X}(\rF,\rref)$ absorbs all $z$-independent flow
dependence, including the endpoint normalizations of the Wilson line and the
leading short-flow-time matching ratio in the quasi case. Extracting the length
slope $\Delta_m(\rF,\rref)$ independently at each $\rF$ in a first stage before
fitting its $(1/\rF-1/\rref)$ dependence gives a practical advantage: it isolates
the linear divergence at each radius without fitting the finite-flow
corrections across different radii with a single global ansatz. To avoid the
short-distance region where the smoothing spheres at the two ends of the Wilson
line overlap strongly, we restrict the first-stage length fit in physical units
to
\begin{equation}
  |z|>2\max(\rF,\rref).
  \label{eq:physical-z-cut}
\end{equation}

\begin{figure}[t]
  \centering
  \includegraphics[width=0.99\textwidth]{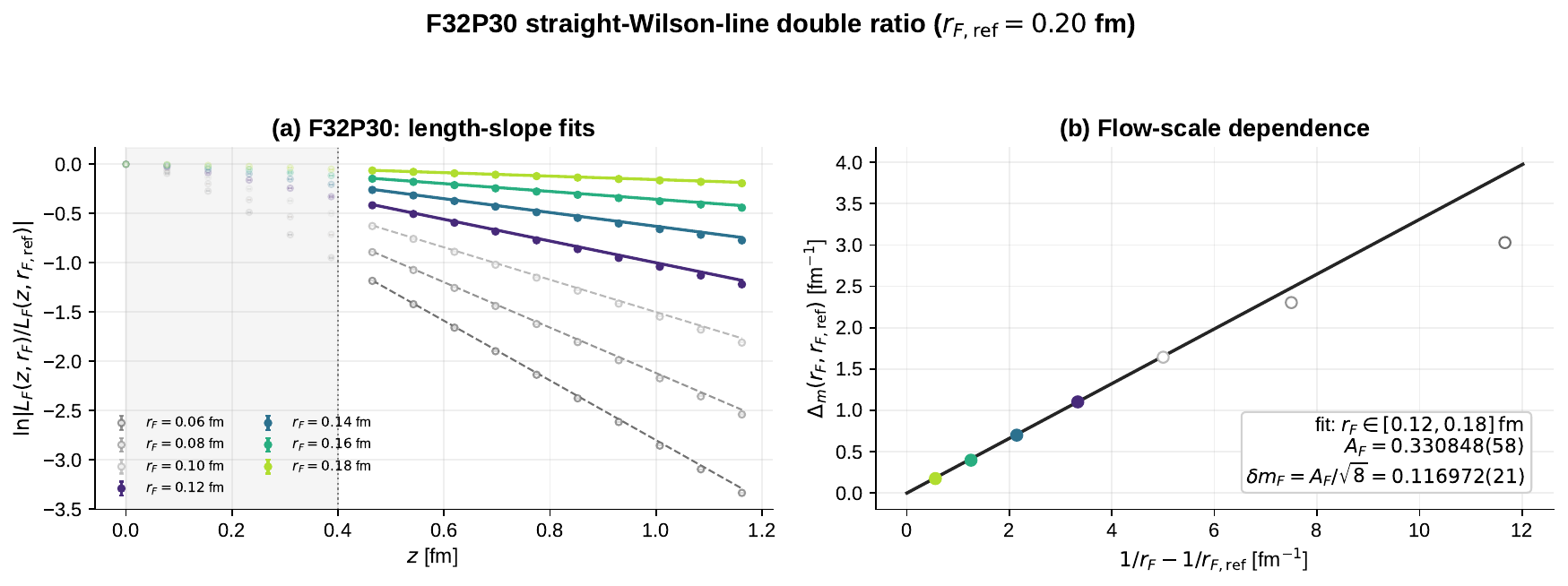}
  \caption{Two-stage F32P30 flow double ratio for the Landau-gauge straight
  Wilson line ($\rref=0.20\,\mathrm{fm}$). (a) First-stage length-slope fits of
  $\ln|\Rcal_L|$ versus $z$ at each flow radius $\rF$ (excluding $\rref$), with
  points in the endpoint-overlap region $z\le2\rref=0.40\,\mathrm{fm}$ shown
  faintly and excluded from the fits. (b) Second-stage fit of the extracted
  slopes $\Delta_m(\rF,\rref)$ versus $1/\rF-1/\rref$. In both panels, colored
  results at $\rF=0.12$--$0.18\,\mathrm{fm}$ enter the central analysis, whereas
  gray series at smaller radii are diagnostic only.}
  \label{fig:wl-double-ratio}
\end{figure}

Fig.~\ref{fig:wl-double-ratio} shows both stages of this analysis for the
Landau-gauge straight Wilson line on F32P30. Above the short-distance cut
$z>2\rref=0.40\,\mathrm{fm}$, $\ln|\Rcal_L|$ is approximately linear in $z$ at every
flow radius, as shown in Fig.~\ref{fig:wl-double-ratio}(a). The extracted slope
$\Delta_m(\rF,\rref)$ follows the expected linear scaling with
$(1/\rF-1/\rref)$ across $\rF=0.12$--$0.18\,\mathrm{fm}$, as shown in
Fig.~\ref{fig:wl-double-ratio}(b). At smaller flow radii, shown as open gray points at $\rF\le 0.10\,\mathrm{fm}$, the slopes begin to
deviate from the linear behavior as $\rF$ approaches the lattice spacing $a$,
and these points are excluded from the fit to reduce $\order(a^2/\rF^2)$ cutoff
effects. Applying the same physical-unit procedure to both ensembles gives
\begin{align}
  A_F^{\rm C24P29}&=0.329408(50),
  &\delta m_F^{\rm C24P29}&=0.116463(18),\nonumber\\
  A_F^{\rm F32P30}&=0.330848(58),
  &\delta m_F^{\rm F32P30}&=0.116972(21).
  \label{eq:straight-line-results}
\end{align}
The central values on the two ensembles differ by only $0.4\%$ despite the
change in lattice spacing from $0.105\,\mathrm{fm}$ to $0.077\,\mathrm{fm}$,
indicating that the leading self-energy behavior is governed primarily by the
physical flow scale.

To check the color-representation dependence, we compare the straight Wilson
line in the fundamental representation ($R=F$) above with the same analysis in
the adjoint representation ($R=A$). At one-loop order in perturbation theory,
the Wilson-line self energy is proportional to the quadratic Casimir, predicting
$A_A/A_F=C_A/C_F=9/4=2.25$~\cite{Monahan:2017hpu,Brambilla:2023vwm}. Our
numerical analysis on the two ensembles gives
\begin{align}
  A_A^{\rm C24P29}&=0.746420(153),\nonumber\\
  A_A^{\rm F32P30}&=0.745289(181).
  \label{eq:adjoint-line-results}
\end{align}
The corresponding adjoint-to-fundamental ratios are
\begin{align}
  (A_A/A_F)^{\rm C24P29}&=2.26594(34),\nonumber\\
  (A_A/A_F)^{\rm F32P30}&=2.25267(39).
  \label{eq:casimir-results}
\end{align}
The central values differ from the one-loop Casimir ratio $9/4$ by
approximately $0.71\%$ for C24P29 and $0.12\%$ for F32P30.

\subsection{Self energy from quasi matrix elements}
\label{subsec:quasi-self-energy}

To test whether the ultraviolet self-energy coefficient $A_F$ determined from
pure-gauge observables also describes the straight Wilson line in hadronic quasi
matrix elements across different channels and momenta, we apply the same flow
double-ratio analysis directly to the hadronic matrix elements.

If one tried to extract the self energy from a quasi matrix element using a
direct joint fit across different flow radii at each $z$,
\begin{equation}
  \ln|\Mcal_X(z,P_z,\rF)|
  =b_X(z,P_z)-A_X\frac{|z|}{\rF},
  \label{eq:ordinary-joint}
\end{equation}
the fit would need an independent intercept $b_X(z,P_z)$ at every separation $z$
to absorb the unknown physical $z$ shape, and would also need to account for
the flow-dependent perturbative matching factor. The resulting $A_X$ can then be sensitive to the
chosen $z$ window and the perturbative matching order.

The flow double ratio in Eq.~\eqref{eq:quasi-double-ratio} avoids this
entanglement by changing the order of operations: dividing $\Mcal_X(z,P_z,\rF)$
by $\Mcal_X(z,P_z,\rref)$ at the same $z$ first cancels the common hadronic
matrix element $\Mcal_X^{\MSbar}(z,P_z,\mu)$ and residual mass $m_0$, and then
determines the ultraviolet self energy from the linear slope in $|z|$. The
$z$-independent ratio of matching factors enters only into the intercept
$B_X(\rF,\rref)$
in Eq.~\eqref{eq:double-ratio-linear-z} and leaves the extracted slope
unchanged.

\begin{figure}[t]
  \centering
  \includegraphics[width=0.99\textwidth]{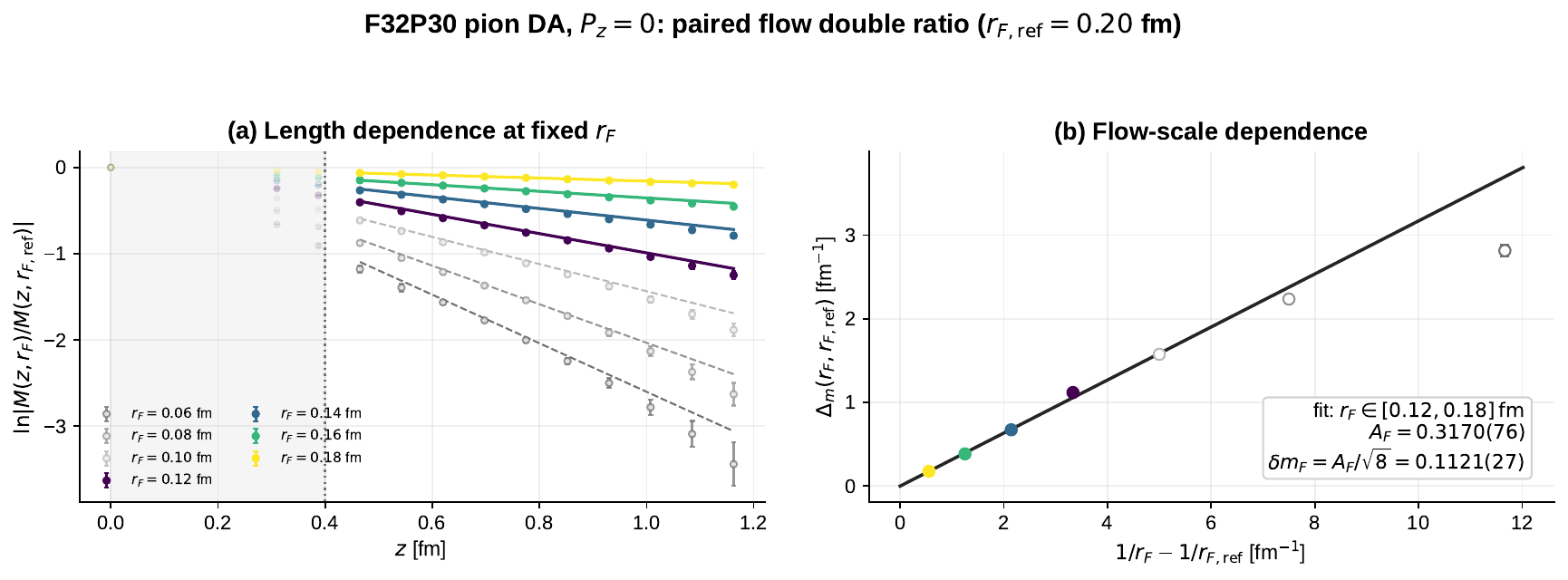}
  \caption{Two stages of the flow double-ratio analysis for the F32P30 pion
  quasi-DA at $P_z=0$ ($\rref=0.20\,\mathrm{fm}$). (a) Linear fits of
  $\ln|\Rcal_X|$ versus $z$ at each flow radius $\rF$ (excluding the reference
  radius $\rref$); faint points at $z\le2\rref=0.40\,\mathrm{fm}$ are excluded.
  (b) Extracted slopes $\Delta_m(\rF,\rref)$ versus $1/\rF-1/\rref$. In both
  panels, colored results at $\rF=0.12$--$0.18\,\mathrm{fm}$ enter the central
  analysis, whereas gray series at smaller radii are diagnostic only.}
  \label{fig:double-ratio-example}
\end{figure}

Fig.~\ref{fig:double-ratio-example} shows this two-stage extraction for the
pion quasi-DA on F32P30 ($a=0.0775\,\mathrm{fm}$) at $P_z=0$. Following the
same two-stage fitting procedure over $z>2\rref=0.40\,\mathrm{fm}$ and
$\rF=0.12$--$0.18\,\mathrm{fm}$ as in Sec.~\ref{subsec:double-ratio}, we
obtain
\begin{equation}
  A_F^{\pi\mathrm{DA}}=0.3170(76),
  \qquad
  \delta m_F^{\pi\mathrm{DA}}=0.1121(27)
  \qquad(P_z=0).
  \label{eq:pion-da-double-ratio-result}
\end{equation}
We also apply the same extraction at boosted momenta $P_z=0.50$ and
$1.00\,\mathrm{GeV}$, as well as to the pion and proton quasi-PDFs; these
results are compared with the pure-gauge determinations in
Sec.~\ref{subsec:self-energy-comparison}.

\subsection{Comparison of different self-energy determinations}
\label{subsec:self-energy-comparison}

\begin{figure}[t]
  \centering
  \includegraphics[width=0.82\textwidth]{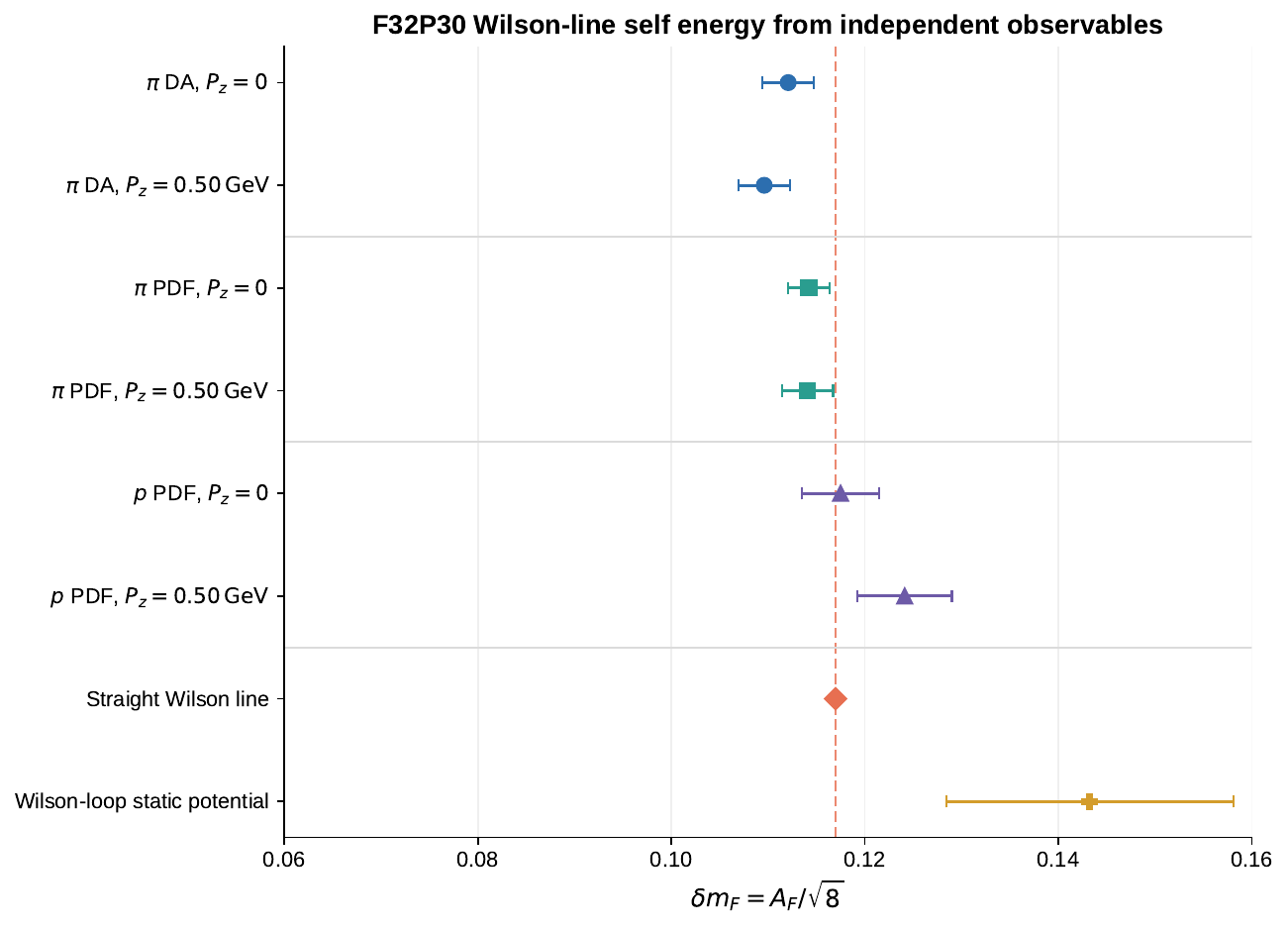}
  \caption{Self-energy coefficient $\delta m_F=A_F/\sqrt8$ extracted from
  independent observables on F32P30. For visual clarity, the three quasi
  channels are shown at $P_z=0$ and $0.50\,\mathrm{GeV}$ (with
  $1.00\,\mathrm{GeV}$ listed in Table~\ref{tab:self-energy-summary}). The
  dashed line marks the Landau-gauge straight Wilson-line determination.
  The static-potential error bar includes the combined jackknife
  statistical and AIC-weighted fit-selection uncertainty.}
  \label{fig:self-energy-summary}
\end{figure}

In the preceding subsections, we have extracted the Wilson-line self-energy
coefficient through three independent approaches: the gauge-invariant static
potential from closed rectangular Wilson loops in
Sec.~\ref{subsec:static-potential}, the flow double ratio of Landau-gauge
straight Wilson lines in Secs.~\ref{subsec:gauge-fixed-line} and
\ref{subsec:double-ratio}, and the flow double ratio of hadronic quasi matrix
elements in Sec.~\ref{subsec:quasi-self-energy}.
Table~\ref{tab:self-energy-summary} and Fig.~\ref{fig:self-energy-summary}
collect and compare the numerical results from all three determinations.
The quoted uncertainties include jackknife statistical errors and,
for the hadronic and static-potential extractions, fit-window systematic
uncertainties estimated using AIC.
The six low-momentum quasi determinations cluster around the straight
Wilson-line result, $\delta m_F^{\rm WL}=0.116972(21)$, as shown in
Fig.~\ref{fig:self-energy-summary}, with their central values differing from it
by no more than about $6\%$. At the highest momentum considered,
$P_z=1.00\,\mathrm{GeV}$, some channel-dependent shifts are visible.
Finite-flow corrections, including $\order(\tau P_z^2)$ terms, cutoff effects,
and dependence on the temporal fit window associated with residual
excited-state contamination may contribute to these differences and require
further assessment.

The static-potential determination, $\delta m_F^V=0.143(15)$,
is about $22\%$ higher than the straight Wilson-line result.
For Wilson loops, the limited signal at large temporal separations $T$
restricts the range over which ground-state dominance can be tested with
the present statistics. The two-state fits and AIC averaging account for
variations among the available temporal fit windows, but these windows
may share a residual excited-state bias that is not fully reflected in
the AIC-based uncertainty. Such contamination could contribute to the
larger self-energy extracted from the static potential. Higher statistics
and fits extending to larger $T$ are needed to assess this effect.

Taken together, these comparisons provide a qualitative numerical check
of the expected universality of the leading Wilson-line self-energy.
At the present level of precision and systematic control, the different
determinations broadly support a common self-energy contribution for the same
straight Wilson line across observables and hadronic states. This supports
using a statistically precise Wilson-line determination as a common
nonperturbative input to the power subtraction in GFRR.

\begin{table}[t]
  \centering
  \caption{Low-momentum F32P30 quasi distribution results and the two pure-gauge
  determinations in the positive convention $\delta m=A/\sqrt8$. Errors include
  jackknife statistical uncertainties and, for the hadronic and
  static-potential determinations, fit-window systematic uncertainties
  estimated using AIC. The
  Landau-gauge straight Wilson-line result is used as the nonperturbative
  subtraction input in Sec.~\ref{sec:step-by-step}.}
  \label{tab:self-energy-summary}
  \begin{tabular}{llcc}
    \toprule
    ensemble & observable & $A$ & $\delta m$ \\
    \midrule
    F32P30 & pion DA, $P_z=0$ & $0.3170(76)$ & $0.1121(27)$ \\
    F32P30 & pion DA, $P_z=0.50\,\mathrm{GeV}$ & $0.3100(76)$ & $0.1096(27)$ \\
    F32P30 & pion DA, $P_z=1.00\,\mathrm{GeV}$ & $0.288(22)$ & $0.1019(78)$ \\
    F32P30 & pion PDF, $P_z=0$ & $0.3230(61)$ & $0.1142(22)$ \\
    F32P30 & pion PDF, $P_z=0.50\,\mathrm{GeV}$ & $0.3227(74)$ & $0.1141(26)$ \\
    F32P30 & pion PDF, $P_z=1.00\,\mathrm{GeV}$ & $0.384(17)$ & $0.1358(60)$ \\
    F32P30 & proton PDF, $P_z=0$ & $0.332(11)$ & $0.1175(40)$ \\
    F32P30 & proton PDF, $P_z=0.50\,\mathrm{GeV}$ & $0.351(14)$ & $0.1241(49)$ \\
    F32P30 & proton PDF, $P_z=1.00\,\mathrm{GeV}$ & $0.251(22)$ & $0.0886(79)$ \\
    \midrule
    C24P29 & straight Wilson line & $0.329408(50)$ & $0.116463(18)$ \\
    F32P30 & straight Wilson line & $0.330848(58)$ & $0.116972(21)$ \\
    F32P30 & static potential, AIC workflow & $0.405(42)$ & $0.143(15)$ \\
    \bottomrule
  \end{tabular}
\end{table}

\section{Numerical implementation of GFRR}
\label{sec:step-by-step}

As formulated in Sec.~\ref{subsec:flow-quasi-renormalization}, gradient flow
introduces a physical smoothing radius $\rF$ that can be varied independently
of the lattice spacing, providing an intermediate continuum-like scheme in
which the ultraviolet divergences of nonlocal quasi operators can be separated
and removed systematically. Combining flowed quasi matrix elements evaluated at
several flow times with the independently determined Wilson-line self-energy
coefficient $A_F^{\rm WL}$ on the same ensembles
(Sec.~\ref{sec:self-energy}) enables their conversion to the $\MSbar$ scheme through
the GFRR framework of Eq.~\eqref{eq:framework-renormalization-map}. The resulting
matrix elements are incorporated into a hybrid scheme for subsequent LaMET
matching. 
Since the ultraviolet renormalization factors are independent of the
external hadronic state, we illustrate this coordinate-space renormalization
using the pion quasi-DA on F32P30 as a demonstration across 
$P_z\in[0,2.50]\,\mathrm{GeV}$, where the statistical signal is cleanest.

\subsection{Normalization at \texorpdfstring{$z=0$}{z=0} and power subtraction}
\label{subsec:local-norm-subtraction}

As introduced in Sec.~\ref{subsubsec:flowed-hadronic-renormalization}, we
work with flowed matrix elements normalized by their local values at $z=0$,
Eqs.~\eqref{eq:da-local-normalization} and \eqref{eq:local-normalization}.
For the pion quasi-DA, this normalization arises directly from the two-point
correlation-function reduction: the ratio of nonlocal and local two-point
correlation functions cancels the common pion-source overlap and yields the
matrix-element ratio
$H_\pi^{\rm DA}(z,P_z,\rF,a)/H_\pi^{\rm DA}(0,P_z,\rF,a)$, while the local
axial-current matrix element in the denominator supplies the normalization
associated with the pion decay constant $f_\pi$ required by the DA definition,
giving $\Mcal_\pi^{\rm DA}(0,P_z,\rF,a)=1$. Evaluating the numerator and
denominator at the same flow time also cancels the common flowed-quark
wave-function factor $\mathring Z_\chi$ in Eq.~\eqref{eq:ringed_Z}, avoiding
its explicit evaluation. For the pion and proton quasi-PDFs, the analogous
$z=0$ normalization places the nonlocal renormalization of all three channels
on the same footing.

For the pion quasi-DA figures below, we remove the conventional phase associated
with symmetric momentum sharing between the quark and antiquark and display
the real part of
\[
  \widetilde{\Mcal}_\pi^{\rm DA}(z,P_z)
  =e^{-iP_z z/2}\Mcal_\pi^{\rm DA}(z,P_z),
\]
where $P_z$ denotes the physical momentum. This phase rotation leaves the
renormalization factors unchanged while making their common action on the
different momentum channels more clearly visible.

\begin{figure}[t]
  \centering
  \includegraphics[width=0.96\textwidth]{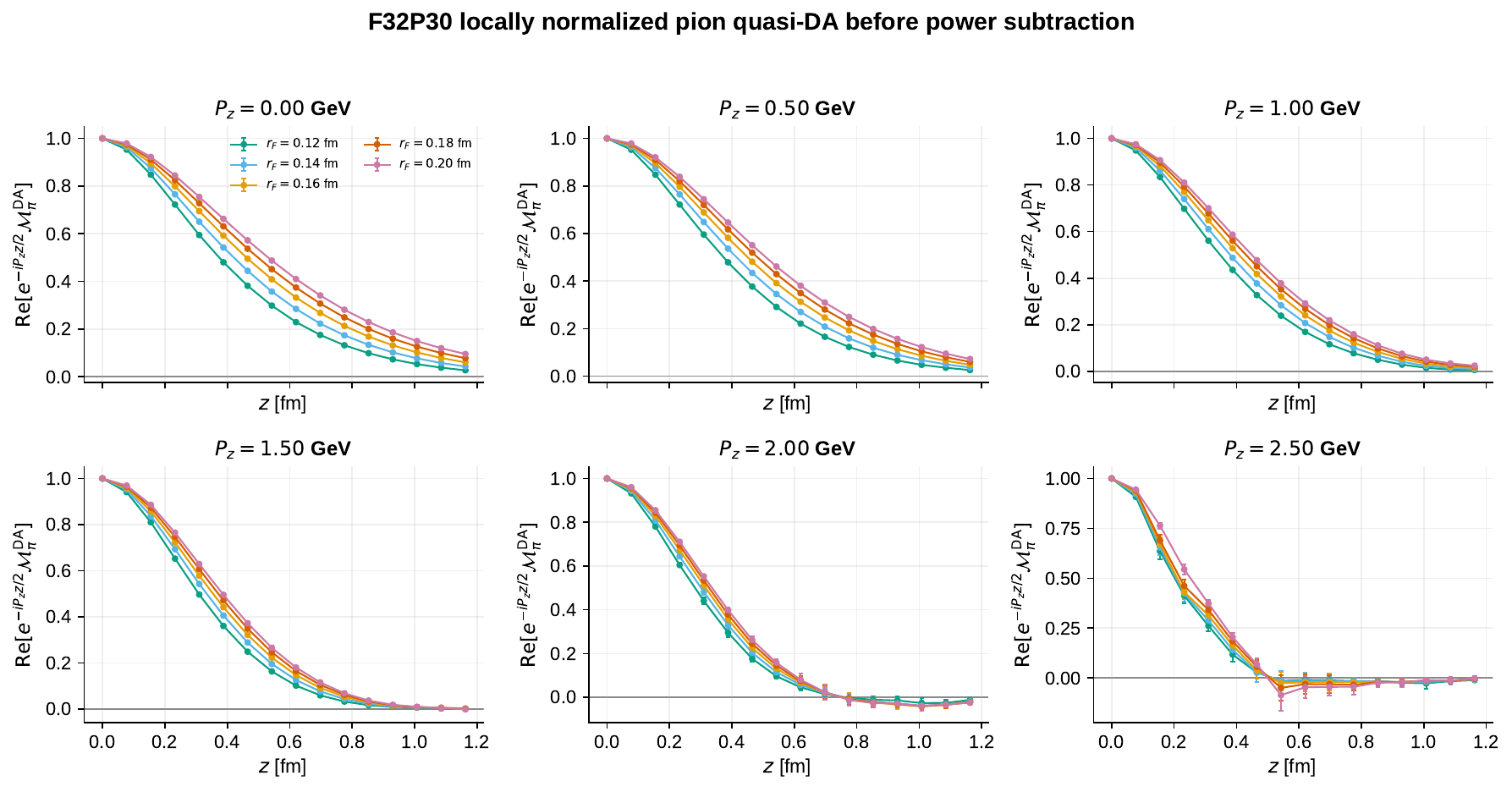}
  \caption{Real part of the phase-rotated F32P30 pion quasi-DA matrix elements
  $\widetilde{\Mcal}_\pi^{\rm DA}(z,P_z,\rF,a)$, defined above and normalized
  at $z=0$ before Wilson-line
  power subtraction across $P_z=0$--$2.50\,\mathrm{GeV}$ and
  $\rF=0.12$--$0.20\,\mathrm{fm}$. Normalization at $z=0$ fixes the origin to unity. At nonzero separation,
  the leading Wilson-line factor $\exp[-A_F|z|/\rF]$ produces stronger
  self-energy suppression at smaller flow radii.}
  \label{fig:bare-flowed-quasi}
\end{figure}

At nonzero spatial separation, the normalized matrix element carries
the length-dependent Wilson-line self energy as well as the logarithmic
matching difference between the nonlocal operator and the local current. As shown in Fig.~\ref{fig:bare-flowed-quasi} for the F32P30 pion quasi-DA,
all curves equal unity at $z=0$. At nonzero separation, the factor
$\exp[-A_F|z|/\rF]$ gives stronger Wilson-line self-energy suppression
at smaller flow radii across all six momenta.

The independent determination of $A_F^{\rm WL}$ from the Landau-gauge
straight Wilson line, Eq.~\eqref{eq:straight-line-results} and
Table~\ref{tab:self-energy-summary}, fixes the linear ultraviolet
subtraction without fitting the hadronic $z$ dependence. We define the
power-subtracted matrix element by
\begin{equation}
  \Mcal_X^{\rm sub}(z,P_z,\rF,a)
  =\exp\!\left(+\frac{A_F^{\rm WL}}{\rF}|z|\right)
   \Mcal_X(z,P_z,\rF,a),
  \label{eq:self-energy-subtraction}
\end{equation}
with the positive sign fixed by the convention in
Eqs.~\eqref{eq:double-ratio-linear-z} and \eqref{eq:flow-law}.

\begin{figure}[t]
  \centering
  \includegraphics[width=0.96\textwidth]{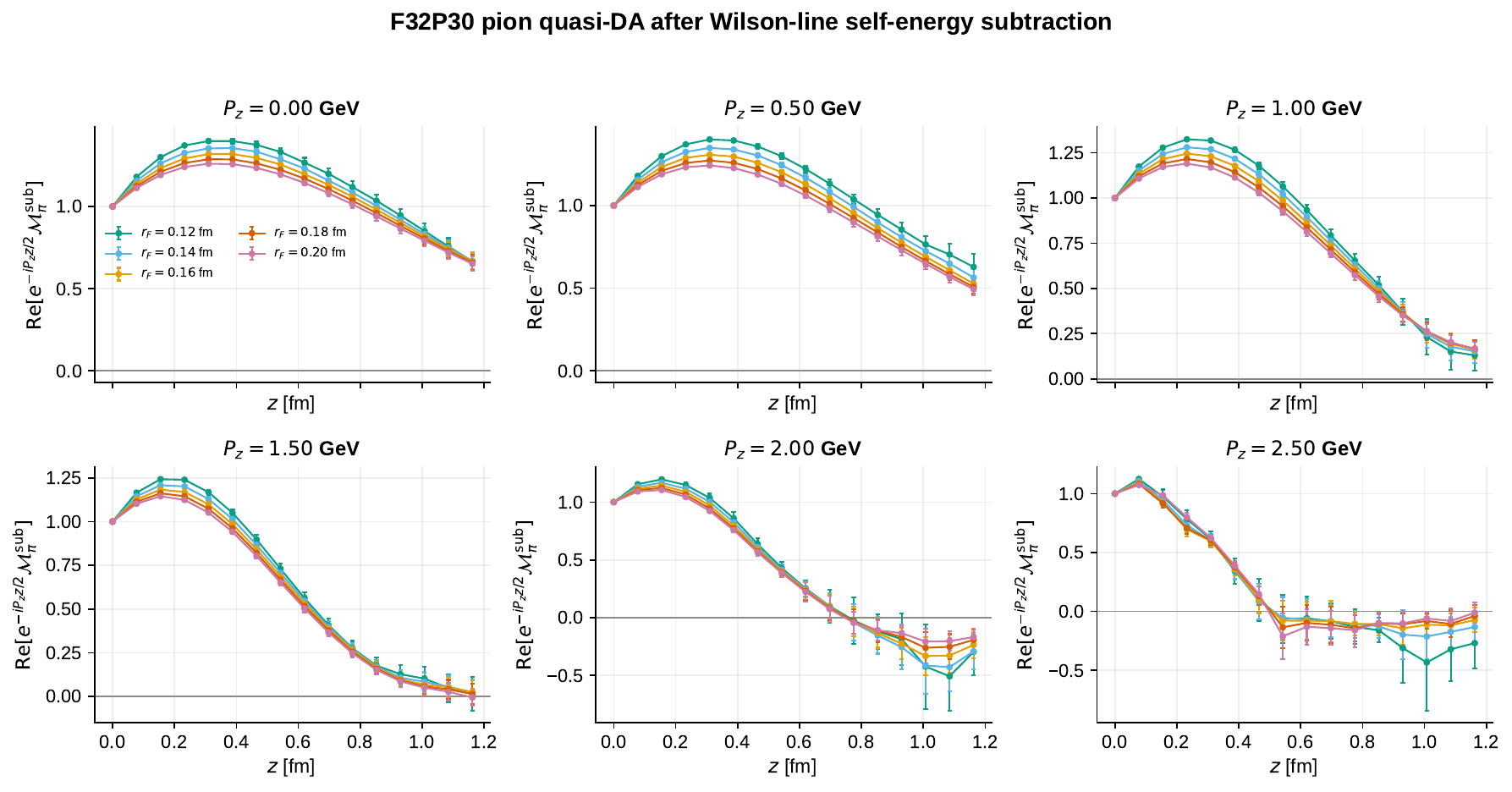}
  \caption{Real part of the phase-rotated F32P30 pion quasi-DA matrix elements
  $\widetilde{\Mcal}_\pi^{\rm sub}(z,P_z,\rF,a)$
  after normalization at $z=0$ and nonperturbative power subtraction using the
  Landau-gauge straight Wilson-line coefficient $A_F^{\rm WL}$ from
  Eq.~\eqref{eq:straight-line-results}. Compared with
  Fig.~\ref{fig:bare-flowed-quasi}, the strong exponential dependence on $\rF$
  is removed across all momenta.}
  \label{fig:self-subtraction}
\end{figure}

Fig.~\ref{fig:self-subtraction} displays the F32P30 pion quasi-DA after the
subtraction in Eq.~\eqref{eq:self-energy-subtraction}. Applying the same
$A_F^{\rm WL}$ removes the dominant exponential separation among flow radii
across all six momenta, bringing the phase-rotated curves for
$\rF\in[0.12,0.20]\,\mathrm{fm}$ into a narrow band and revealing the
momentum dependence of the boosted matrix element. The
remaining flow-radius dependence contains the logarithmic
flow-to-$\MSbar$ matching factor and finite-flow corrections, treated
through perturbative conversion and zero-flow-time extrapolation below.

\subsection{Continuum limit at fixed flow radius}

At positive flow time, gradient flow suppresses ultraviolet modes above $1/\rF$.
Once QCD is renormalized and the common flowed-quark wave-function factor is
canceled in the $z=0$-normalized ratios of
Sec.~\ref{subsec:local-norm-subtraction}, the resulting flowed matrix
elements admit a finite continuum limit~\cite{Luscher:2011bx,Luscher:2013cpa}.
As indicated in Eq.~\eqref{eq:framework-renormalization-map}, the continuum
limit $a\to 0$ is defined at fixed physical separation $z$, momentum $P_z$, and
flow radius $\rF>0$,
\begin{equation}
  \Mcal_X^{\rm sub}(z,P_z,\rF)
  =\lim_{a\to0}\Mcal_X^{\rm sub}(z,P_z,\rF,a),
  \label{eq:continuum-limit}
\end{equation}
before taking the zero-flow-time limit $\tau\to 0$. 
The order of these limits is essential and cannot be interchanged. 
Reversing this order at finite $a$ would drive the smoothing radius below the lattice spacing
($\rF\lesssim a$), returning the ultraviolet regulator to the lattice cutoff
and losing the $\order(a^2/\rF^2)$ organization of discretization effects.

Within our flow window $\rF\in[0.12,0.20]\,\mathrm{fm}$, the normalized
matrix elements on C24P29 and F32P30 agree closely at matched physical flow
radii, as shown in Fig.~\ref{fig:uv-scale-transfer} and the extracted 
coefficients $A_F^{\rm WL}$ differ by only $0.4\%$ in central value.
This weak observed lattice-spacing dependence motivates us to proceed with
the subsequent renormalization steps directly on the two ensembles available,
without performing a continuum extrapolation in the present work. Matching
and flow-time fits are carried out separately on each ensemble, and the
lattice-spacing argument $a$ is suppressed below. We use F32P30 to illustrate
these steps; C24P29 is included in the residual-mass determination and the
final comparison in Sec.~\ref{subsec:hybrid-result}.

\subsection{Flow-to-\texorpdfstring{$\MSbar$}{MSbar} matching}
\label{subsec:flow-to-msbar}

The nonlocal operator and the local current in the normalized matrix element
have different ultraviolet matching coefficients, $C_q(\tau,\mu)$ and
$C_{\rm loc}(\tau,\mu)$. Their ratio $C_{\rm R}(\tau,\mu)$, defined in
Eq.~\eqref{eq:gf-normalized-matching}, contains the logarithmic flow-time
dependence $\ln(2\mu^2\tau)$. This coefficient converts the power-subtracted
matrix elements at different flow radii to the $\MSbar$ scheme at a common
renormalization scale.

As formulated in Sec.~\ref{subsubsec:flowed-operator-matching}, the
RG-improved coefficient $C_{\rm R}^{\rm RG}(\tau,\mu)$ in
Eq.~\eqref{eq:gf-rg-matching} matches at the natural flow scale
$\mu_\tau=e^{-\gamma_E/2}/\sqrt{2\tau}=2e^{-\gamma_E/2}/\rF$ and evolves to
the target scale $\mu=2\,\mathrm{GeV}$. The conversion at fixed nonzero separation is
\begin{equation}
  \Mcal_X^{\MSbar}(z,P_z,\tau,\mu)
  =\frac{\Mcal_X^{\rm sub}(z,P_z,\rF)}
        {C_{\rm R}^{\rm RG}(\tau,\mu)},
  \label{eq:flow-ms-conversion}
\end{equation}
where $\Mcal_X^{\MSbar}(z,P_z,\tau,\mu)$ denotes the quasi matrix element
converted to the $\MSbar$ scheme.
At $z=0$, the numerator and denominator of the local-normalization
ratio coincide, and the exactly normalized value
$\Mcal_X^{\MSbar}(0,P_z,\tau,\mu)=1$ is retained.
In evaluating $C_{\rm R}^{\rm RG}(\tau,\mu)$,
we use fixed-$N_f=3$ one-loop running and one-loop flow-to-$\MSbar$ matching
with $\alpha_s(2\,\mathrm{GeV})=0.2960$. At leading order in the short-flow-time expansion, the conversion
rescales the data at nonzero separations by the same factor
$1/C_{\rm R}^{\rm RG}(\tau,\mu)$ for a given flow radius.

\begin{figure}[t]
  \centering
  \includegraphics[width=0.66\textwidth]{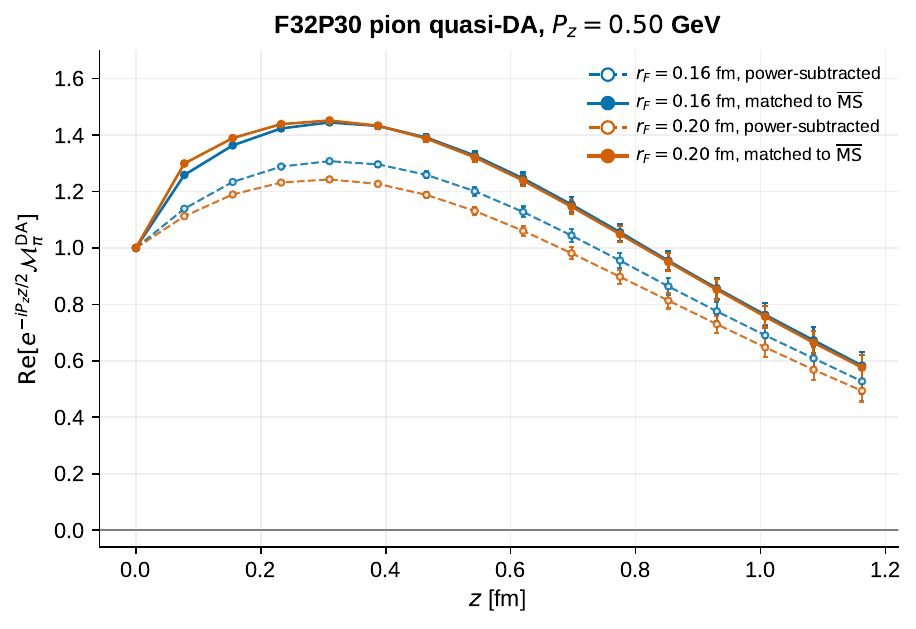}
  \caption{Flow-to-$\MSbar$ conversion of the phase-rotated F32P30 pion quasi-DA at
  $P_z=0.50\,\mathrm{GeV}$ for $\rF=0.16\,\mathrm{fm}$ (blue) and
  $0.20\,\mathrm{fm}$ (orange). Dividing the power-subtracted matrix elements at nonzero separation
  (open dashed curves) by $C_{\rm R}^{\rm RG}(\tau,\mu)$
  at $\mu=2\,\mathrm{GeV}$ gives the $\MSbar$-matched curves (filled solid),
  bringing the two flow radii into closer agreement.}
  \label{fig:matching}
\end{figure}

\begin{figure}[t]
  \centering
  \includegraphics[width=0.96\textwidth]{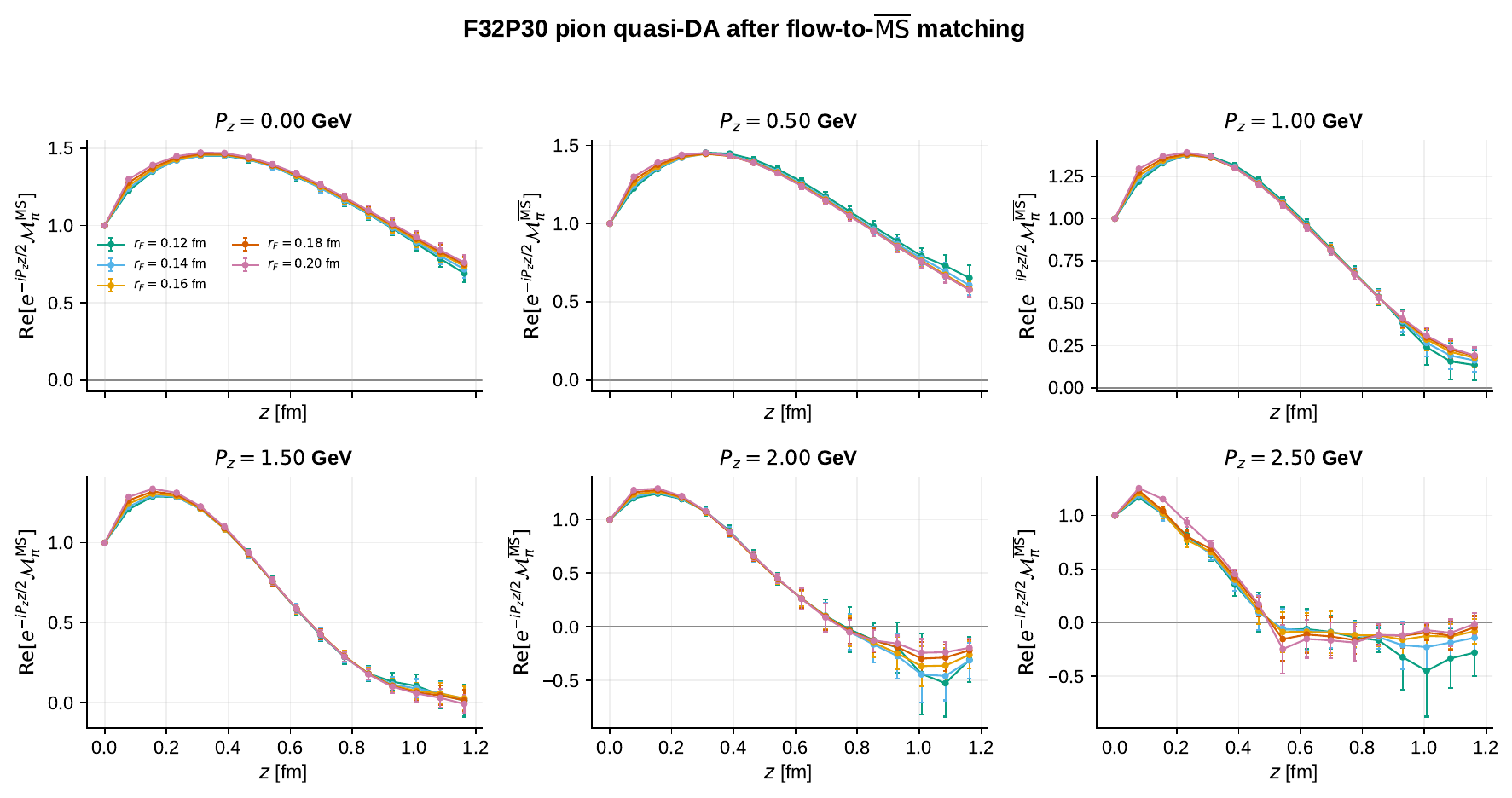}
  \caption{Real part of the phase-rotated, RG-improved flow-to-$\MSbar$
  matched F32P30 pion quasi-DA matrix elements at $\mu=2\,\mathrm{GeV}$ over
  the full momentum range for $\rF=0.12$--$0.20\,\mathrm{fm}$.}
  \label{fig:matching-overview}
\end{figure}

Fig.~\ref{fig:matching} illustrates the conversion for the F32P30 pion
quasi-DA at $P_z=0.50\,\mathrm{GeV}$. The matched curves for
$\rF=0.16$ and $0.20\,\mathrm{fm}$ are brought into closer alignment in the
phase-rotated real part. Fig.~\ref{fig:matching-overview} shows
the matched matrix elements at the common scale $\mu=2\,\mathrm{GeV}$
for all six momenta and five flow radii, which provide the input for the
zero-flow-time extrapolation below.

\subsection{Zero-flow-time limit}

At positive flow time, the short-flow-time expansion in
Eq.~\eqref{eq:GF_MS_matching_full} contains higher-dimensional operator
contributions suppressed by powers of $\tau$, arising from the finite smoothing
radius $\rF=\sqrt{8\tau}$~\cite{Luscher:2011bx,Brambilla:2023vwm}. Removing
these finite-flow effects requires taking the zero-flow-time limit at fixed
nonzero separation $z$, formally defined after the continuum limit as
\begin{equation}
  \Mcal_X^{(0)}(z,P_z,\mu)
  =\lim_{\tau\to0}\Mcal_X^{\MSbar}(z,P_z,\tau,\mu),
  \label{eq:tau-zero-limit}
\end{equation}
in accordance with Eq.~\eqref{eq:framework-renormalization-map}. In the
finite-lattice-spacing analysis below, we continue to denote the extrapolated
estimates by $\Mcal_X^{(0)}(z,P_z,\mu)$.

Motivated by the leading $\order(\tau)$ correction in the short-flow-time
expansion, we perform correlated linear fits in $\tau$ over the five radii
$\rF\in[0.12,0.20]\,\mathrm{fm}$ on paired jackknife samples with a regularized
covariance matrix. This linear extrapolation is carried out at all measured
nonzero separations $z$, including the short-distance points subsequently used
for the residual-mass fit in Sec.~\ref{subsec:residual-mass}, yielding
zero-flow-time intercept estimates across the full spatial range.

\begin{figure}[t]
  \centering
  \includegraphics[width=0.99\textwidth]{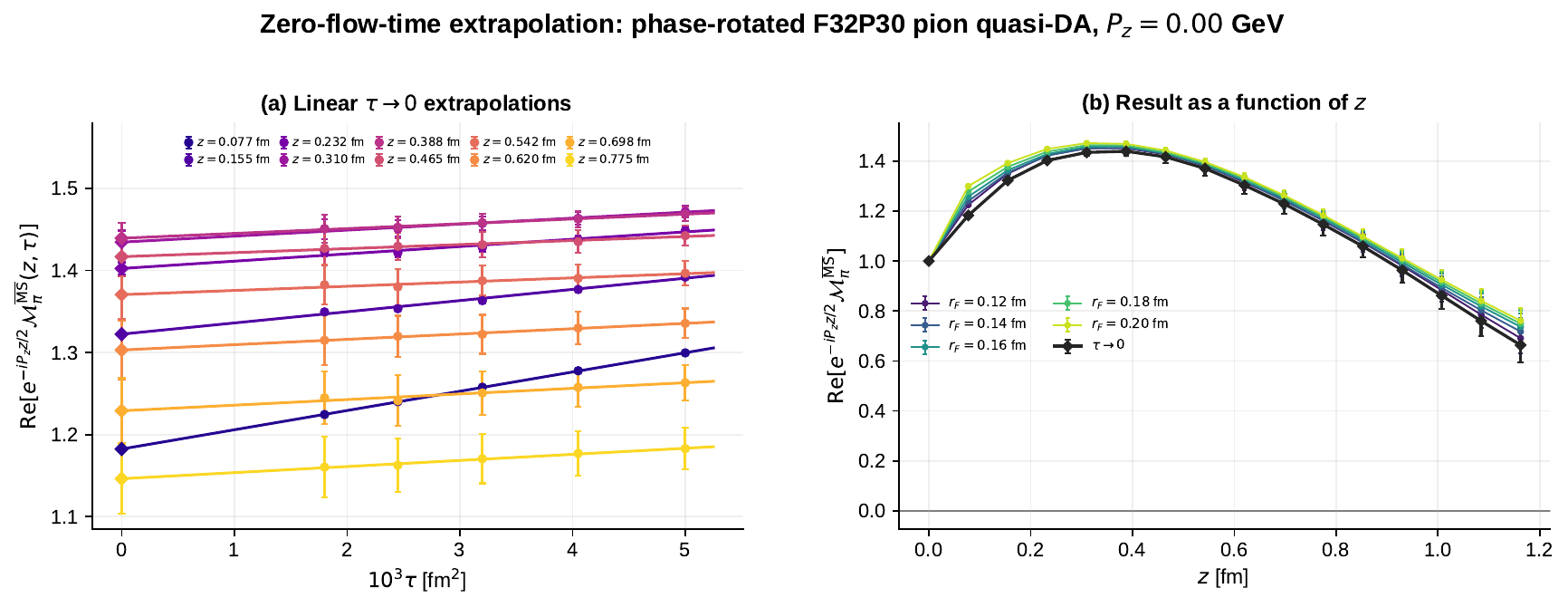}
  \caption{Zero-flow-time extrapolation of the real part of the phase-rotated,
  RG-matched F32P30 pion quasi-DA at $P_z=0$. The phase rotation is trivial at
  zero momentum. (a) Correlated fixed-$z$ linear fits in
  $\tau$ for representative separations $z/a=1,\ldots,10$, with the intercepts
  at $\tau=0$ marked by diamonds. (b) The extrapolated $\tau\to 0$ profile
  (black, representing $\Mcal_\pi^{(0)}$) compared with the five finite-flow
  curves $\rF=0.12$--$0.20\,\mathrm{fm}$.}
  \label{fig:tau-extrapolation-pz0}
\end{figure}

\begin{figure}[t]
  \centering
  \includegraphics[width=0.99\textwidth]{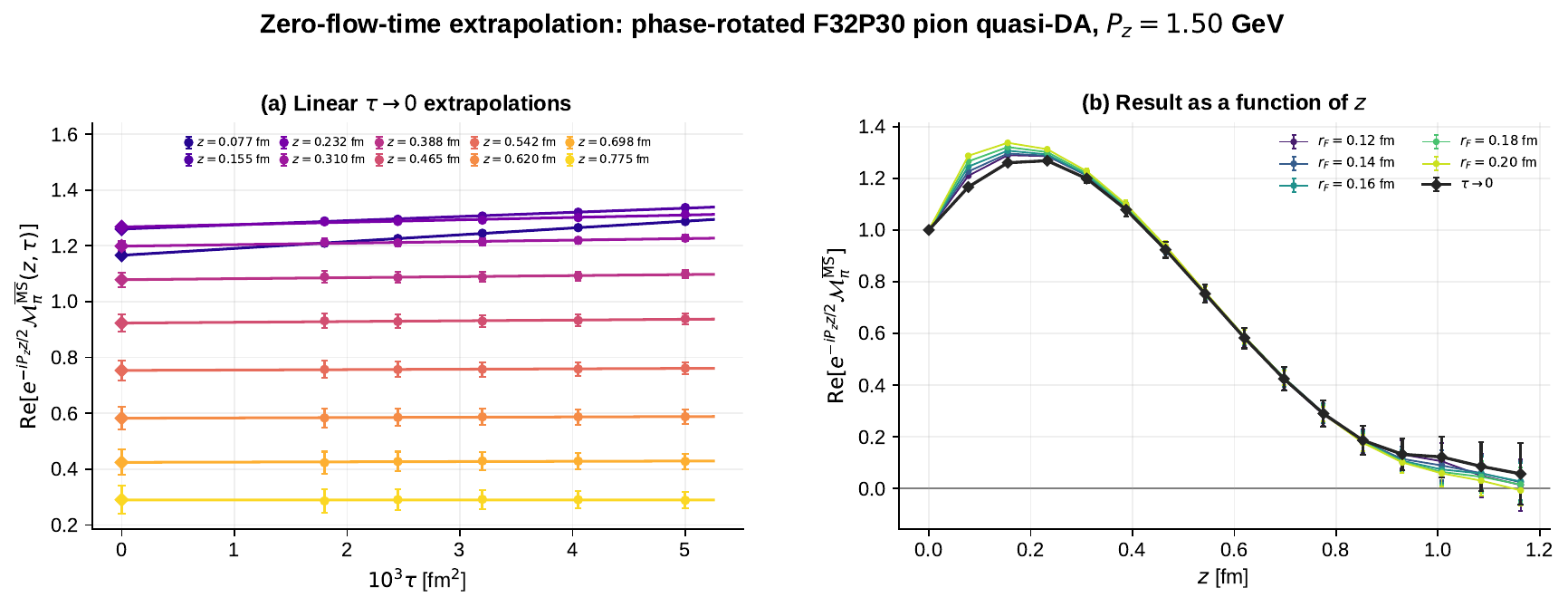}
  \caption{Zero-flow-time extrapolation of the real part of the phase-rotated,
  RG-matched F32P30 pion quasi-DA at $P_z\simeq1.50\,\mathrm{GeV}$, following the same layout as
  Fig.~\ref{fig:tau-extrapolation-pz0}.}
  \label{fig:tau-extrapolation-pz3}
\end{figure}

\FloatBarrier

Figs.~\ref{fig:tau-extrapolation-pz0} and \ref{fig:tau-extrapolation-pz3}
show representative $\tau\to0$ extrapolations for the F32P30 pion quasi-DA
at $P_z=0$ and $P_z\simeq1.50\,\mathrm{GeV}$. In the left panels, the
matched matrix elements at fixed $z$ exhibit an approximately linear
dependence on $\tau$ over $\rF=0.12$--$0.20\,\mathrm{fm}$, supporting the
linear fits within the chosen flow window. The black curves in the right
panels show the resulting zero-flow-time estimates
$\Mcal_\pi^{(0)}(z,P_z,\mu)$. These estimates retain the flow-independent
residual-mass factor $e^{-m_{0,F}|z|}$ in
Eq.~\eqref{eq:framework-renormalization-map}, which cannot be separated by
varying the flow time and requires a short-distance matching condition.

\subsection{Residual-mass prescription from short-distance matching}
\label{subsec:residual-mass}
The flow double ratio allows us to determine and subtract the
linear divergence, but leaves the residual mass $m_0$ undetermined.
We extract $m_0$ by matching the zero-momentum matrix element
$\Mcal_X^{(0)}(z,0,\mu)$ to its short-distance OPE,
with both renormalization group resummation (RGR) and
leading-renormalon resummation (LRR)~\cite{Holligan:2023rex}.
The central Wilson coefficient is evaluated at the natural scale
$\mu_z=2e^{-\gamma_E}/|z|$ and evolved to $\mu$.
The initial scale is varied coherently as
$\mu_{\mathrm i}=c\mu_z$, with $c\in[0.75,1.5]$.
The expressions below correspond to $c=1$; for scale variations,
$\mu_z$ is replaced by $\mu_{\mathrm i}$ throughout the matching
coefficient and its RG evolution.
At NLO, the resulting Wilson coefficient reads
\begin{equation}
  C_{\rm LRR}(z,\mu)
  = U(\mu,\mu_z)
  \left(
    C_0^{\rm NLO}(z,\mu_z)
    + |z|\mu_z
    \left[
      S_{\rm PV}(\alpha_s(\mu_z))
      - r_0\alpha_s(\mu_z)
    \right]
  \right),
  \label{eq:c0-lrr}
\end{equation}
where
\begin{align}
  C_0^{\rm NLO}(z,\mu_z)
  &= 1+\frac{\alpha_s(\mu_z)C_F}{2\pi}
  \left[
    \frac{3}{2}\ln\frac{\mu_z^2z^2e^{2\gamma_E}}{4}
    +\frac{5}{2}
  \right], \\
  S_{\rm PV}(\alpha_s)
  &= \frac{4\pi N_m}{\beta_0}
    \mathrm{PV}\int_0^\infty \mathrm{d}u\,
    \frac{e^{-4\pi u/(\beta_0\alpha_s)}}
         {(1-2u)^{1+b}}
    \left[1+c_1(1-2u)+c_2(1-2u)^2\right].
  \label{eq:lrr-pv}\\
    U(\mu,\mu_z)
  &= \exp\left[
\int_{\alpha_s(\mu_z)}^{\alpha_s(\mu)}
    \mathrm{d}\alpha\,
    \frac{\gamma(\alpha)}{\beta(\alpha)}
  \right]
  = \left[
    \frac{\alpha_s(\mu_z)}{\alpha_s(\mu)}
  \right]^{3C_F/\beta_0}
\end{align}
The symbol $\mathrm{PV}$ denotes the principal-value prescription for the
Borel integral.
Here $b=\beta_1/(2\beta_0^2)$, with
$\beta_1=102-38N_f/3$,
and $r_0=N_m(1+c_1+c_2)$.
We use $N_m=0.575$, $c_1=-0.16382$, and
$c_2=0.23717$ for $N_f=3$.

The residual mass $m_0$ is extracted from the short-distance behavior
\begin{equation}
  \ln|\Mcal_X^{(0)}(z,0,\mu)|
  -\ln C_{\rm LRR}(z,\mu)
  =b_0-m_0|z|+\order(z^2\Lambda_{\rm QCD}^2),
  \label{eq:m0-fit}
\end{equation}
where the constant offset $b_0$ accounts for a finite normalization
difference.
The constant $b_0$ is retained as a nuisance intercept in the
short-distance fit, but is not applied as a finite renormalization to the
matrix elements, which have already been normalized at $z=0$. The
residual-mass subtraction is therefore applied at all momenta as
\begin{equation}
  \Mcal_X^{\rm LRR}(z,P_z,\mu)
  =e^{m_0|z|}\Mcal_X^{(0)}(z,P_z,\mu),
  \label{eq:m0-apply}
\end{equation}
which preserves the local normalization
$\Mcal_X^{\rm LRR}(0,P_z,\mu)=1$.

\begin{figure}[t]
  \centering
  \includegraphics[width=0.88\textwidth]{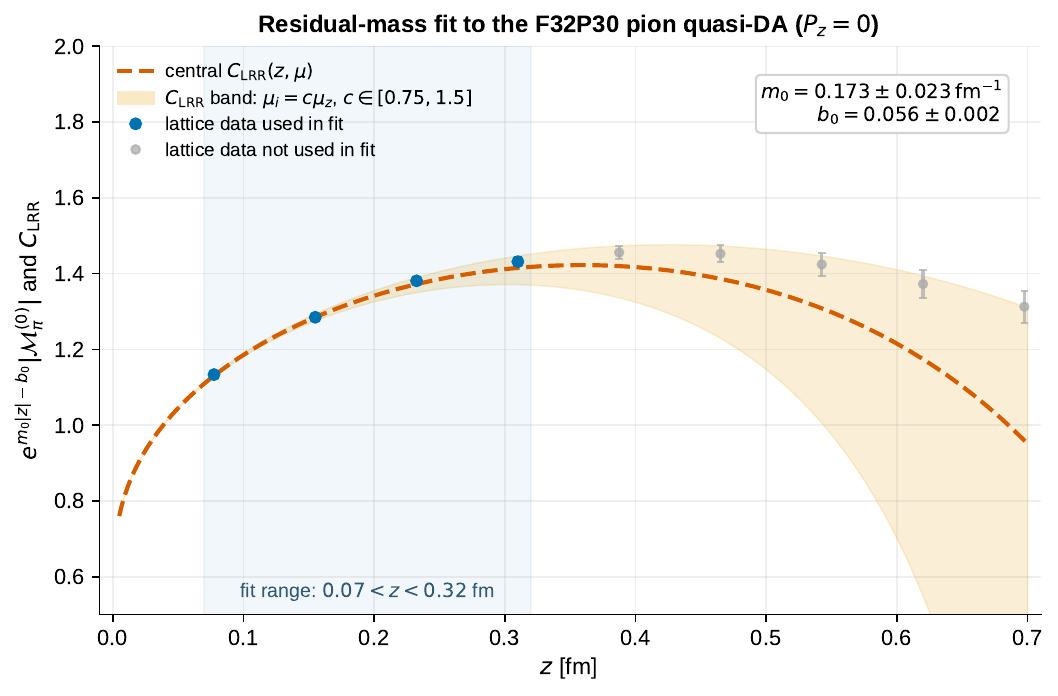}
  \caption{Short-distance determination of the residual mass $m_0$ from the
  F32P30 pion quasi-DA at $P_z=0$. The transformed zero-flow-time matrix
  elements $e^{m_0|z|-b_0}|\Mcal_\pi^{(0)}(z,0,\mu)|$ are shown in blue for
  the points included in the fit and in gray for those outside the fit range.
  The dashed orange curve denotes the central $C_{\rm LRR}(z,\mu)$ at
  $\mu=2\,\mathrm{GeV}$, while the orange band shows the initial-scale
  variation $\mu_{\mathrm{i}}=c\mu_{\mathrm{z}}$, with
  $\mu_{\mathrm{z}}=2e^{-\gamma_{\mathrm{E}}}/|z|$ and
  $c\in[0.75,1.5]$. The residual-mass fit uses the central curve.
  The blue shaded region marks the fit window
  $0.07\,\mathrm{fm}<|z|<0.32\,\mathrm{fm}$.}
  \label{fig:m0-fit}
\end{figure}

Using the zero-flow-time matrix elements obtained in the
previous subsection,
we fit Eq.~\eqref{eq:m0-fit} using the central $C_{\rm LRR}(z,\mu)$ over the window
$0.07\,\mathrm{fm}<|z|<0.32\,\mathrm{fm}$. To stabilize the strongly
correlated short-distance covariance matrix, we employ an SVD cutoff
$\epsilon_{\rm SVD}=10^{-4}$. As
shown in Fig.~\ref{fig:m0-fit} for the F32P30 pion quasi-DA at $P_z=0$,
multiplying by $e^{m_0|z|-b_0}$ aligns the lattice matrix element closely with
$C_{\rm LRR}(z,\mu)$ inside the shaded window. The gray points outside
this window are shown for comparison and are not included in the fit. This fit
yields $m_0=0.173(23)_{\rm stat}\,\mathrm{fm}^{-1}$ on F32P30 and
$m_0=0.171(11)_{\rm stat}\,\mathrm{fm}^{-1}$ on C24P29, fixing the finite
residual-mass prescription on each ensemble. To estimate the perturbative
uncertainty associated with the initial scale, we repeat the complete
correlated fit for each coherent choice $c\in[0.75,1.5]$ and take the envelope
of the resulting $m_0$ values, rather than treating the band as an
uncorrelated point-by-point uncertainty. This gives additional scale
uncertainties ${}^{+0.024}_{-0.009}\,\mathrm{fm}^{-1}$ on F32P30 and
${}^{+0.018}_{-0.002}\,\mathrm{fm}^{-1}$ on C24P29.

\subsection{Hybrid short- and long-distance result}
\label{subsec:hybrid-result}

Although the renormalized matrix element
$\Mcal_X^{\rm LRR}(z,P_z,\mu)$ is now available in the
$\MSbar$ scheme, it is still advantageous to employ the
ratio scheme at short distances for subsequent LaMET matching.
In conventional LaMET calculations, the ratio scheme not only
cancels common multiplicative renormalization factors and
short-distance logarithms $\ln(\mu^2z^2)$ but also suppresses
discretization effects at $|z|\sim a$%
~\cite{Ji:2020brr}.
In the gradient-flow scheme, suppressing lattice cutoff effects also
requires $a\ll\rF$, making the hierarchy $a\ll\rF\ll|z|$ difficult
to maintain at the shortest separations.
A similar advantage is expected for gradient-flow
renormalization: when the hierarchy $\rF\ll |z|$ is not
satisfied, the additional finite-flow corrections to the
matching coefficient are expected to largely cancel in
the ratio.
We therefore define
\begin{equation}
  \Mcal_X^{\rm ratio}(z,P_z)
  =\frac{\Mcal_X^{\rm LRR}(z,P_z,\mu)}
         {\Mcal_X^{\rm LRR}(z,0,\mu)},
  \label{eq:ratio_scheme}
\end{equation}
which cancels these common short-distance logarithms and enforces the ratio
normalization at short distances. At large $|z|$, by contrast, dividing by
$\Mcal_X^{\rm LRR}(z,0,\mu)$ would fold the nonperturbative long-distance shape
of the zero-momentum matrix element into the renormalization factor.

\begin{figure}[t]
  \centering
  \includegraphics[width=0.98\textwidth]{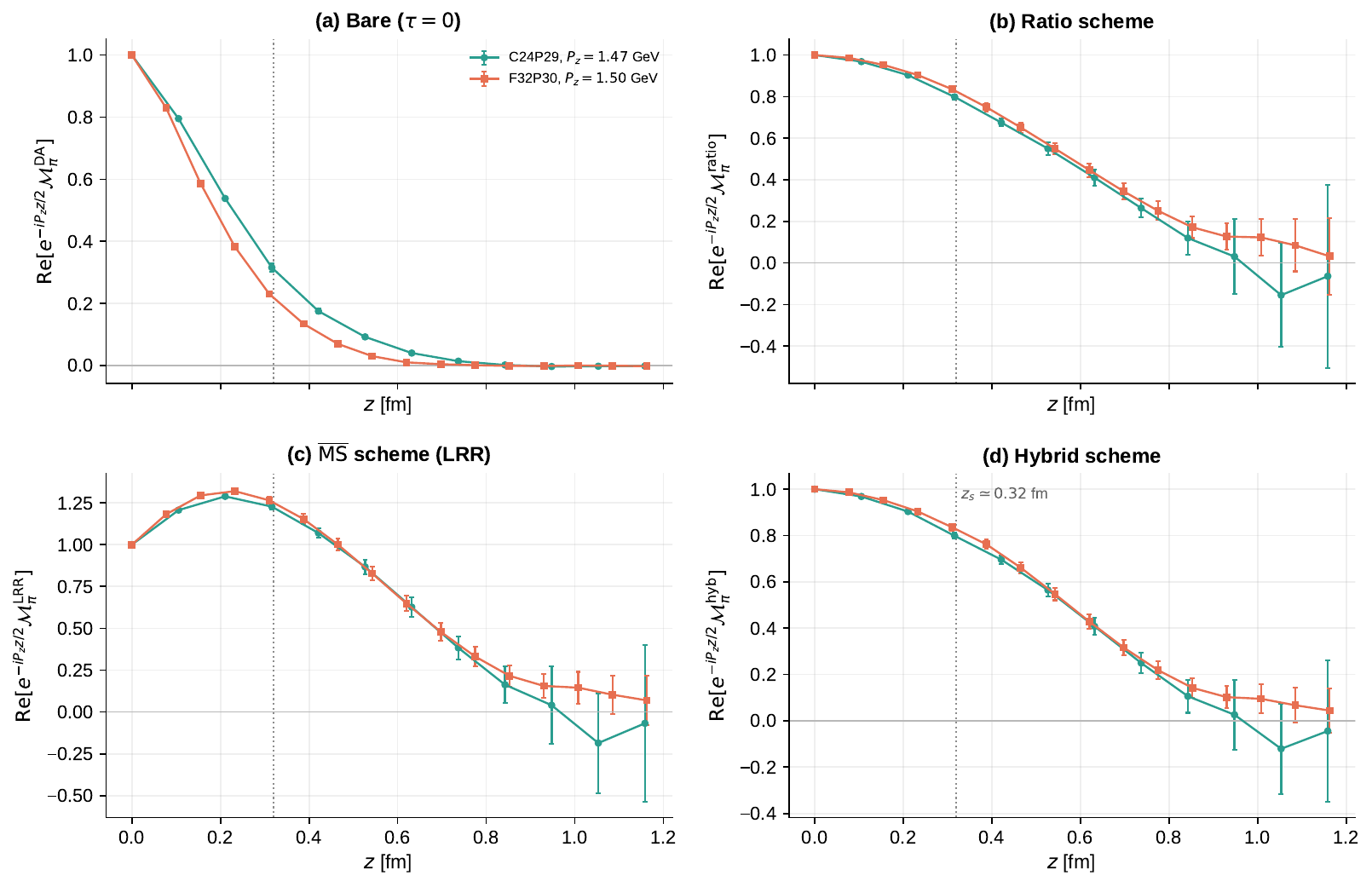}
  \caption{Real parts of the phase-rotated coordinate-space pion quasi-DA
  matrix elements, with the phase convention defined above, on C24P29
  ($P_z=1.47\,\mathrm{GeV}$, teal) and F32P30 ($P_z=1.50\,\mathrm{GeV}$, orange)
  before and after renormalization using GFRR: (a) bare ($\tau=0$), (b) ratio scheme,
  (c) $\MSbar$ scheme with the LRR prescription
  ($\Mcal_\pi^{\rm LRR}$), and (d) hybrid scheme
  ($\Mcal_\pi^{\rm hyb}$). Both ensembles are overlaid in each panel.
  Panel~(c) shows
  $\Mcal_\pi^{\rm LRR}=e^{m_0|z|}\Mcal_\pi^{(0)}$,
  as defined in Eq.~\eqref{eq:m0-apply}, where $\Mcal_\pi^{(0)}$
  is the zero-flow-time $\MSbar$ result. The fit intercept $b_0$
  is not applied as an additional normalization, and the value at
  $z=0$ remains unity.
  The dotted line indicates the switching region
  $z_s\simeq0.32\,\mathrm{fm}$; each ensemble uses the nearest available
  lattice separation.}
  \label{fig:hybrid-result}
\end{figure}

To separate the two regimes, we adopt the hybrid construction
\cite{Ji:2020brr,LatticePartonCollaborationLPC:2025vhd}, retaining the ratio scheme below a
short-distance switch point $z_s$ and fixing the denominator at $z_s$ for
larger separations. On each ensemble, $z_s\simeq0.32\,\mathrm{fm}$ is
chosen as the nearest available lattice separation:
\begin{equation}
  \Mcal_X^{\rm hyb}(z,P_z)=
  \begin{cases}
    \displaystyle
    \frac{\Mcal_X^{\rm LRR}(z,P_z,\mu)}
         {\Mcal_X^{\rm LRR}(z,0,\mu)},
      & |z|\le z_s,\\[1.1em]
    \displaystyle
    \frac{\Mcal_X^{\rm LRR}(z,P_z,\mu)}
         {\Mcal_X^{\rm LRR}(z_s,0,\mu)},
      & |z|>z_s.
  \end{cases}
  \label{eq:hybrid-result}
\end{equation}
The two branches meet continuously at $|z|=z_s$. Since $b_0$ is used only as
a fit intercept and is not applied to $\Mcal_X^{\rm LRR}$, the local
normalization is preserved before either branch is formed. For $|z|>z_s$, the fixed denominator sets
the normalization at the joining point while preserving the $z$ dependence of
$\Mcal_X^{\rm LRR}(z,P_z,\mu)$, so the finite mass from
Sec.~\ref{subsec:residual-mass} enters this branch through $e^{m_0(|z|-z_s)}$.
Fig.~\ref{fig:hybrid-result} compares C24P29 and F32P30 at approximately matched physical
momentum ($P_z=1.47$ and $1.50\,\mathrm{GeV}$) with
$z_s\simeq0.32\,\mathrm{fm}$. In Fig.~\ref{fig:hybrid-result}(a), the bare
nonflowed curves exhibit a large lattice-spacing separation; after applying GFRR
in panel~(c), the two ensembles agree closely across the full
$z$ range. Panel~(d) shows the final hybrid matrix elements.

Eq.~\eqref{eq:hybrid-result} gives the final coordinate-space matrix element
from GFRR, which can subsequently be extrapolated at
large $|z|$, Fourier transformed to momentum space, and matched perturbatively
onto the light-cone distribution using a kernel consistent with this hybrid
definition and the chosen residual-mass prescription.

\FloatBarrier

\section{Conclusion}
\label{sec:conclusion}

We have developed and numerically investigated the GFRR framework
for spatially nonlocal Wilson-line operators on two $N_f=2+1$ clover
ensembles with lattice spacings $a=0.10530(18)$ and $0.07746(18)\,\mathrm{fm}$.
By introducing a physical smoothing radius $\rF=\sqrt{8\tau}$ that can be
varied independently of the lattice spacing, gradient flow replaces the lattice
cutoff in the leading ultraviolet suppression of bare quasi matrix elements and
strongly reduces their lattice-spacing dependence at matched physical $\rF$.
Evaluating nonlocal observables across several flow radii on the same gauge
configurations further enables flow ratios that cancel unknown long-distance
functions and flow-independent residual mass factors, while absorbing
$z$-independent matching factors into a free intercept. The linear Wilson-line
self energy can thus be isolated directly from the $|z|$ slope without modeling
the long-distance spatial dependence of the matrix element.

Analyzing the flow-radius dependence of Landau-gauge straight Wilson lines
yields $\delta m_F=0.116463(18)$ on C24P29 and $0.116972(21)$ on F32P30, whose
central values differ by only $0.4\%$ across the two lattice spacings.
Independent determinations from flow double ratios of low-momentum pion
quasi-DA, pion and proton quasi-PDFs, and from a gauge-invariant Wilson-loop
static-potential analysis yield consistent values within their statistical and
systematic uncertainties, supporting the expected state independence of the
leading Wilson-line self energy across the studied channels and low momenta.
Comparing adjoint and fundamental straight Wilson lines gives
$A_A/A_F=2.26594(34)$ on C24P29 and $2.25267(39)$ on F32P30, differing in
central value from the one-loop Casimir ratio $C_A/C_F=9/4$ by approximately
$0.71\%$ and $0.12\%$, respectively.

Using the high-precision straight-Wilson-line coefficient $A_F^{\rm WL}$ as an
independent nonperturbative input, we demonstrated the GFRR procedure
on the pion quasi-DA across
$P_z\in[0,2.50]\,\mathrm{GeV}$ on both ensembles. With normalization at $z=0$
canceling the common flowed-quark wave-function factor, the independent
Wilson-line subtraction removes the leading linear divergence across all
momenta without fitting the hadronic $z$ dependence. RG-improved perturbative
matching accounts for the logarithmic flow-time dependence and converts the
matrix elements to the $\MSbar$ scheme, and correlated linear $\tau\to 0$ fits
provide finite-lattice-spacing zero-flow-time estimates across all measured
nonzero separations. Short-distance matching of the zero-momentum estimates to
a leading-renormalon-resummed Wilson coefficient fixes the finite residual mass
on each ensemble, and a continuous hybrid construction handles the common
short-distance coordinate-space logarithms $\ln(\mu^2z^2)$ while preserving the
mass-subtracted long-distance dependence for subsequent LaMET matching.
Extending the analysis to additional lattice spacings will enable a controlled
continuum extrapolation at fixed physical $\rF$, while incorporating
higher-order perturbative flow-matching coefficients will improve the
accuracy of the scheme conversion.
Within the present two-ensemble study, these results establish GFRR as
a practical framework for renormalizing pion quasi-DA matrix elements
and point toward its broader application to meson, baryon, and general nonlocal
Wilson-line operators.

\acknowledgments
We thank Xiang-Peng Wang and members of the Lattice Parton Collaboration
for useful discussions. Numerical calculations were performed at the
Southern Nuclear Science Computing Center. This work is supported in part
by the National Natural Science Foundation of China under Grant
Nos.~12275277, 12435004, 12575084, and 12647104.
H.~Y.~Bai is also supported by the National Key Research and Development
Program of China under Grant No.~2023YFA1606000.
J.~Hua is supported by the Natural Science Foundation of Guangdong
Province under Grant No.~2025A151501219 and the Guangdong Special
Support Plan (2025TQ09A655).
The lattice QCD calculations employed the PyQUDA software
package~\cite{Jiang:2024lto} and
QUDA~\cite{Clark:2009wm,Babich:2011np,Clark:2016rdz}.
We have made a comparison with  LQCDMaster~\cite{Gao:2026cpd}, PhysMaster~\cite{Miao:2025sms,Tan:2026ier}, 
and LaMET-Agent~\cite{He:2026kft} in preliminary code validation and  
data analysis during the early stages of this work.

\appendix

\section{Conventions and statistical procedure}
\label{app:conventions}

Our reported coefficient $A_R$ is dimensionless, as seen directly in the
exponent $A_R|z|/\rF$ (and in $\Delta_m(\rF,\rref)|z|$ from
Eq.~\eqref{eq:flow-law}). Since $1/\sqrt\tau=\sqrt8/\rF$, the leading
exponential factor can equivalently be written in terms of $\sqrt\tau$ as
\begin{equation}
  \exp\!\left(-\frac{A_R|z|}{\rF}\right)
  =\exp\!\left(-\frac{\delta m_R|z|}{\sqrt\tau}\right),
  \qquad
  \delta m_R=\frac{A_R}{\sqrt8}>0.
\end{equation}
Here $\delta m_R>0$ is a positive dimensionless coefficient, whereas
$\delta m(\tau)$ in Eq.~\eqref{eq:gf-line-mass} of
Sec.~\ref{subsubsec:flowed-operator-matching} denotes the signed,
dimensionful Wilson-line mass appearing in $e^{\delta m(\tau)|z|}$. For
$\rF<\rref$, the positive coefficient enters the double-ratio logarithm as
$\ln|\Rcal|=B-A_R(1/\rF-1/\rref)|z|$. For a closed rectangular Wilson loop, the
additive constant of the static potential contains the self energies of two
temporal Wilson lines; this factor of two is divided out before comparing with
a single straight Wilson line or a quasi operator.

All matrix elements normalized at $z=0$ and ratios of observables at different
flow radii are constructed on paired configuration-level delete-one jackknife
samples. The resulting covariance matrices preserve the strong statistical
correlations among different spatial separations and flow radii in the
correlated fits. The same paired jackknife procedure is used for the
adjoint-to-fundamental ratio. Stability with respect to covariance-matrix
regularization and fit-window choices has been checked across the three
hadronic quasi channels, where the resulting variations are comparable in
magnitude to the statistical uncertainties.

\bibliographystyle{unsrt}
\bibliography{references}

\end{document}